\documentclass[journal ]{new-aiaa}
\usepackage[utf8]{inputenc}
\usepackage{textcomp}
\usepackage{subcaption}
\usepackage{graphicx}
\usepackage{amsmath}
\usepackage[version=4]{mhchem}
\usepackage{siunitx}
\usepackage{longtable,tabularx}
\usepackage{soul}
\usepackage{xcolor}
\colorlet{editcolor}{red}
\newcommand{\edit}[1]{\textcolor{editcolor}{#1}}

\title{Eclipse Observables for Navigation in Cislunar Space}

\author{Ian M. Down\footnote{Astrodynamics/Satellite Navigation Engineer}
}
\affil{Advanced Space, LLC, Westminster, Colorado, 80234}
\author{James McElreath\footnote{Aerospace Engineer}}
\affil{Space Exploration Engineering, LLC, Laurel, Maryland, 20725}
\author{Manoranjan Majji\footnote{Professor, Department of Aerospace Engineering, AIAA Associate Fellow}}
\affil{Texas A\&M University, College Station, Texas, 77843}

\begin{document}

\maketitle

\begin{abstract}
Autonomous navigation independent of Earth-based tracking remains a challenge for spacecraft operating in cislunar space. This paper proposes onboard eclipse detection as an opportunistic measurement set to supplement the state estimation of cislunar trajectories. Penumbral and umbral entry and exit times detected from standard power telemetry are formulated as implicit zero-measurements on the boundaries of second-order shadow cones computed from planetary ephemerides. These observables are processed in an unscented Kalman filter. The event-time uncertainty that weights each measurement is derived from physical sources excluded in the spherical body model, including telemetry sampling, Earth's atmosphere, and topography/oblateness of the occulting body. The approach is demonstrated both in simulation for an $L_1$ halo orbit and with flight telemetry and orbit determination products from the CAPSTONE spacecraft during the January 2026 lunar eclipse, where umbral contacts timed directly from the power telemetry agree with the geometric contacts to within one sample interval. In both cases, hardware-free eclipse event timing shows utility in uncertainty reduction, especially as the initial solution quality is degraded.
\end{abstract}

\section{Introduction}

\lettrine{A}{utonomous} onboard deep space navigation systems continue to play an important role in the utilization and exploration of the Solar System. By reducing the need for ground-in-the-loop measurement collection, processing, and navigation solution generation, more spacecraft can operate further from Earth under fewer running constraints with less reliance on high-demand ground systems like NASA's Deep Space Network (DSN). This becomes especially important at far distances as cislunar space traffic is expected to increase in the coming decade surrounding humanity's return to the lunar surface. Navigation in this region of space is further complicated by chaotic dynamics found in trajectories orbiting near or around libration points, which feel competing gravitational attraction from both the Earth and the Moon.

Onboard measurement collection and processing for inertial state estimation of trajectories far from Earth comes in a variety of methods \cite{christian2009review}. Optical navigation methods as a sub-category leverage several different measurement types \cite{bradley2020cislunar,christian2026fundamentals}. The limb or centroid of a celestial body such as a planet or moon can be processed from an image and used as a line-of-sight measurement. Similarly, these image features can be compared against background stars to determine an angular separation measurement between the spacecraft and celestial body. If a spacecraft can obtain clear enough images of a celestial body's surface, surface features may be extracted and referenced with respect to a database for more accurate line-of-sight and relative bearing angle measurements. More recently, cross-range and Doppler measurements between NASA's Lunar Reconnaissance Orbiter and Advanced Space's CAPSTONE spacecraft have demonstrated how cooperative inter-spacecraft measurements can be used for absolute inertial navigation in cislunar space \cite{cheetham2021cislunar,thompson2022cislunar}. Another class of autonomous navigation technologies proposes the collection X-ray pulsar signals because of their consistent generation in the cosmic background \cite{sheikh2006spacecraft}. This list of autonomous deep space navigation methods is not meant to be exhaustive; many variations of the methods mentioned above have been developed in literature.

This paper proposes the use of onboard eclipse detection as opportunistic measurements to be processed in a navigation filter for uncertainty reduction of cislunar trajectories \cite{mcelreath2025eclipse}. Generally speaking, the extreme thermal fluctuation and loss of power generation that occur while a deep space spacecraft passes through an eclipse warrant an avoidance of this phenomenon by design. In fact, libration point orbit design in the Earth-Moon system often centers around the Sun-Earth-Moon synodic period so that a majority of eclipses can be avoided given the proper phasing \cite{zimovan2020near}. NASA's Gateway space station was planned to reside in a 9:2 synodic resonant southern near rectilinear halo orbit (NRHO) to avoid long duration Earth eclipses, but planned to encounter numerous lunar eclipses during its operation \cite{williams2017targeting}. Lunar-resonant astrophysics missions similarly incorporate eclipse mitigation into their trajectory design, as demonstrated by TESS's 2:1 resonant orbit \cite{dichmann2014tess}, and the planned Lazuli Space Observatory \cite{mcelreath2026lazuli}. These trajectories were designed to minimize eclipse durations but still experience or plan to experience many eclipses over their mission lifetimes. Actively avoiding eclipse events require fuel expenditure, and ultimately decreases mission lifespan for non-refuel-able spacecraft. Instead, it is common for spacecraft to meet maximum eclipse survival constraints by hardware design, as was the case for the CAPSTONE spacecraft, which flew the same NRHO planned for Gateway \cite{cheetham2021cislunar}.

Detecting the times at which eclipse events occur can be done with existing onboard sensors such as thermocouples, thermistors, and various current/voltage sensors monitoring power generation used in nominal operations. This effectively makes this navigation augmentation ``free" in the sense that it is then software-defined. Connecting these times to the respective Sun-Earth-Moon geometry from ephemeris data results in a subsequent measurement of the spacecraft's state. If a spacecraft passes directly through an eclipse (both penumbral and umbral cones), four measurements can be retrieved through this sensor data which can then be used to reduce spacecraft state uncertainty. Those measurements include penumbral entry, umbral entry, umbral exit, and penumbral exit. Earlier demonstrations of this principle in perturbed, two-body Earth orbits were presented by Fitzgerald~\cite{fitzgerald2021autonomous} and expanded upon by DeWalch and Fitzgerald~\cite{dewalch2024enhancing}, who used an unscented Kalman filter (UKF) to process times of penumbral and umbral entry/exit. Those studies confirmed the viability of relying solely on ambient sunlight measurements to maintain a useful orbit solution when eclipses were frequent, especially for small satellites with limited hardware. A similar approach explored the use of illumination-derived data profiles to resolve unique observability characteristics in multi-vehicle tracking filters \cite{stoker2023aiding}. These works relied on the two-body conic framework, and took advantage of repeatable and frequent lighting conditions for spacecraft in low-Earth and low-Mars orbit. The work presented in this paper generalizes the concept of using eclipse transition time observables to non-conic cislunar trajectories where eclipse events are sparse. While this measurement type cannot be used as the primary navigation solution in the cislunar regime because of the event's sparsity, it acts as a supplementary uncertainty reduction mechanism when available, virtually for free. This has the potential to delay or cancel a subsequent primary measurement collection activity like celestial body imaging, possibly allowing more time for primary data collection and/or provide a geometrically diverse observable in a batch-style state update.

The paper is outlined as follows. Section \ref{sec:2} reviews a Sun-Earth-Moon ephemeris model, while Sec. \ref{sec:3} details the eclipse measurement models used for penumbra/umbra entry/exit, as well as effects of non-spherical celestial bodies on eclipse geometry. Section \ref{sec:4} discusses how these measurement models are incorporated into a navigation filter, and how measurement uncertainty is derived from the physical sources of eclipse event time modeling error. Section \ref{sec:5} demonstrates the numerical performance of the proposed framework, with both a simulated example, and real telemetry and state solutions from the CAPSTONE spacecraft. Lastly, Sec. \ref{sec:6} concludes the paper.

\section{The Sun-Earth-Moon System}\label{sec:2}

The circular restricted three-body problem (CR3BP) is the most fundamental coherent model of cislunar space. Sun geometry is included in this model without dynamic perturbation by simply borrowing its coplanar time explicit expression from the bi-circular restricted four-body problem (BCR4BP). This combined model was used in the authors' previous work to first demonstrate eclipse detection as a measurement model in cislunar space \cite{mcelreath2025eclipse}. In addition to the full BCR4BP, other dynamical models like the Hill restricted four-body problem (HR4BP) and its quasi-periodic, elliptical, and inclined variations all model the Sun-Earth-Moon system to differing levels of realism compared to an ephemeris model \cite{park2025bridging,park2025frequency}. These intermediate dynamical models serve as important stepping stones to successfully transition multi-year ballistic cislunar libration point trajectories to high fidelity ephemeris equivalents while retaining desired geometric and stability properties \cite{sanaga2026systematic}. Gravitational parameter $\mu$, mean distance from the central body $\bar{a}$, and body radius $R$ are given in Table \ref{tab:sem_parameters}, where $\mu$ and $\bar{a}$ are taken from the DE440 ephemeris (with $\bar{a}$ averaged over the year 2026) and $R$ is taken from the NAIF \texttt{pck00011} planetary constants kernel, using the equatorial radius for Earth.
\begin{table}[hbt!]
\caption{\label{tab:sem_parameters} Sun-Earth-Moon Parameters}
\centering
\begin{tabular}{lccc}
\hline
Body    & $\mu$ [km$^3$/s$^2$]& $\bar{a}$ [km]& $R$ [km] \\\hline
Sun     & 1.3271244$\times10^{11}$ & --                  & 6.9570$\times10^5$\\
Earth   & 3.9860044$\times10^5$    & 1.49620$\times10^8$ & 6.3781$\times10^3$\\
Moon    & 4.9028001$\times10^3$    & 3.84638$\times10^5$ & 1.7374$\times10^3$\\
\hline
\end{tabular}
\end{table}

To test eclipse measurement models in a navigation filter,  a periodic libration point orbit from the Earth-Moon CR3BP is transitioned directly to an ephemeris model via multiple-shooting. The standard point mass ephemeris model for a spacecraft under the influence of $N$ gravitational bodies in an inertial frame is given as
\begin{equation}
    \ddot{\boldsymbol{r}} = -\frac{\mu_j}{r^3}\boldsymbol{r} + \sum_{i=1,\,i\neq j}^N \mu_i\left(\frac{\boldsymbol{r}_i - \boldsymbol{r}}{||\boldsymbol{r}_i - \boldsymbol{r}||^3} - \frac{\boldsymbol{r}_i}{r_i^3}\right)\label{eq:ephem}
\end{equation}
where subscript $j$ denotes the primary body about which the inertial frame is centered. It is conventional for cislunar trajectories to use either the Earth or the Moon as the primary body depending on which is the dominating acceleration seen throughout the time horizon considered. Selecting the correct primary body results in more precise numerical propagation. To propagate a spacecraft's trajectory in the ephemeris model, Solar System ephemerides are queried at the corresponding epoch from the NASA NAIF SPICE database at each integration time step. In this work, this is done using the DE440 kernel via the MICE interface for \textsc{Matlab}.

\section{Eclipse Measurement Modeling}\label{sec:3}

\begin{figure}[htbp!]
     \centering
     \begin{subfigure}{0.7\textwidth}
         \centering
         \includegraphics[width=0.9\textwidth]{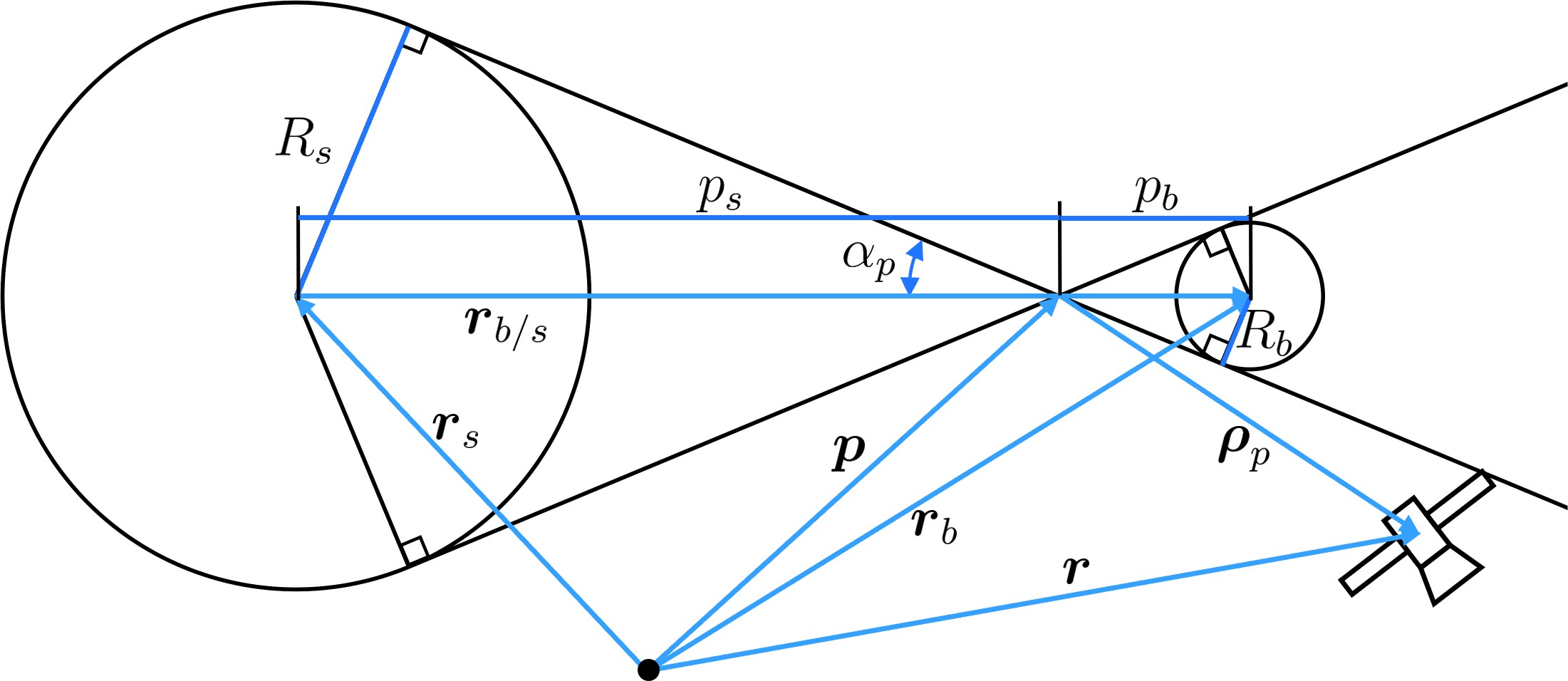}
         \caption{Penumbral Cone}
     \end{subfigure}\\
     \begin{subfigure}{0.7\textwidth}
         \centering
         \includegraphics[width=0.9\textwidth]{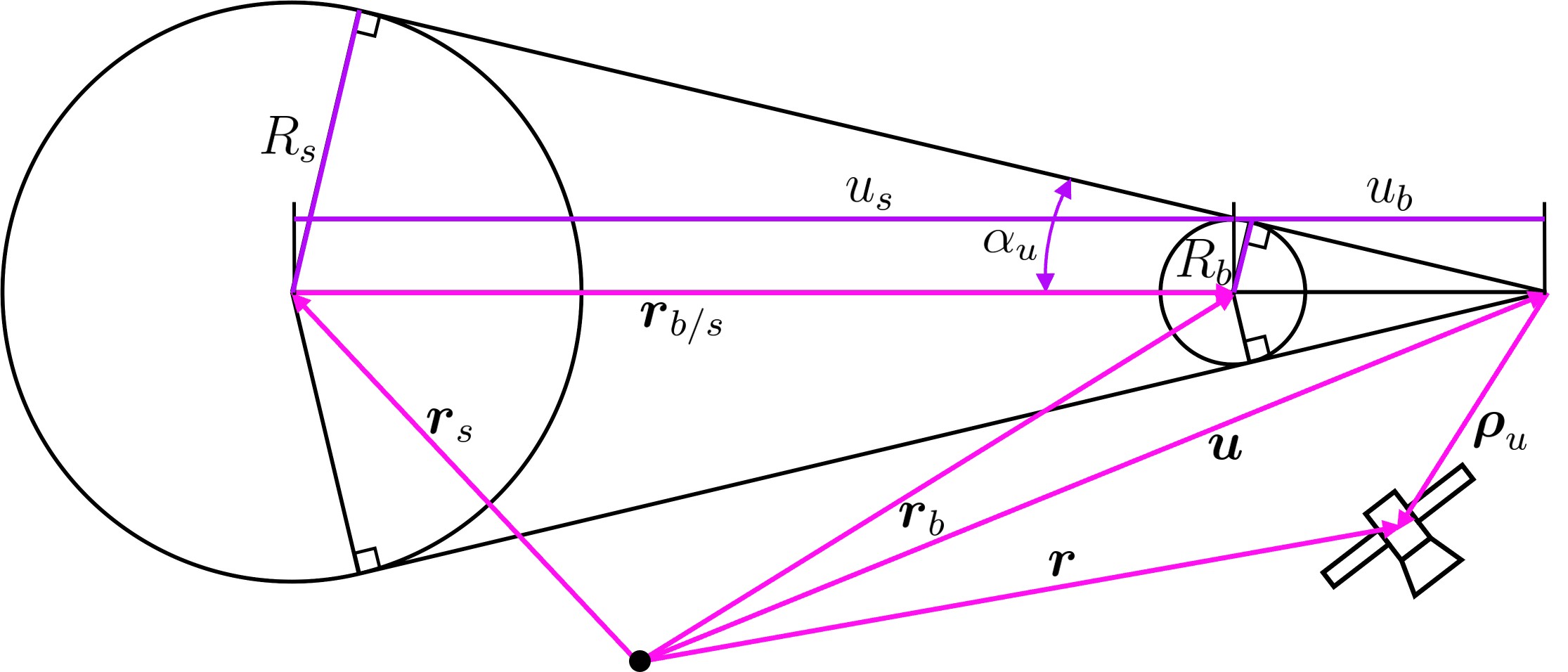}
         \caption{Umbral Cone}
     \end{subfigure}
     \caption{Diagrams of penumbral and umbral eclipse geometry for spherical emitting and occluding bodies.}
     \label{fig:eclipse_diagrams}
\end{figure}

The occlusion of a spherical secondary body (Earth or Moon) with respect to a primary spherical emitting body (Sun) creates three spatial cones: penumbra, umbra, and antumbra. The penumbral and umbral cone geometries are depicted in Fig. \ref{fig:eclipse_diagrams}, with corresponding equations in Table \ref{tab:eclipse_eqs}. An observer inside the penumbral cone experiences a varying fraction of total light, while an observer inside the umbral cone experiences no light from that source at all. The particular distances and angles associated with each cone for the Sun-Earth-Moon system are given in Table \ref{tab:sem_eclipse} using the radii from Table \ref{tab:sem_parameters} and the mean Sun-Earth and Sun-Moon distances from DE440 averaged over the year 2026, $1.49620\times10^8$ km and $1.49633\times10^8$ km, respectively.
\begin{table}[hbt!]
\caption{\label{tab:eclipse_eqs} Computation of Eclipse Geometry for Spherical Bodies}
\centering
\renewcommand{\arraystretch}{2} 
\begin{tabular}{|c|c|}
    \hline
    \textbf{Penumbra} & \textbf{Umbra} \\
    \hline
    $\displaystyle \sin\alpha_p=\frac{R_s}{p_s}=\frac{R_b}{p_b}$ & $\displaystyle \sin\alpha_u=\frac{R_s}{u_s}=\frac{R_b}{u_b}$ \\
    
    $\displaystyle p_s+p_b=||\boldsymbol{r}_{b/s}||$ & 
    $\displaystyle u_s-u_b=||\boldsymbol{r}_{b/s}||$ \\

    $\displaystyle p_s=\frac{||\boldsymbol{r}_{b/s}||}{1+R_b/R_s}\quad \text{and}\quad p_b=\frac{||\boldsymbol{r}_{b/s}||}{1+R_s/R_b}$ & 
    $\displaystyle u_s=||\boldsymbol{r}_{b/s}||+u_b\quad \text{and}\quad u_b=\frac{||\boldsymbol{r}_{b/s}||}{R_s/R_b-1}$ \\

    $\displaystyle \boldsymbol{p}=\boldsymbol{r}_s+p_s\hat{\boldsymbol{r}}_{b/s}=\boldsymbol{r}_s + \frac{1}{1+R_b/R_s}\boldsymbol{r}_{b/s}$ & 
    $\displaystyle \boldsymbol{u}=\boldsymbol{r}_b+u_b\hat{\boldsymbol{r}}_{b/s}=\boldsymbol{r}_b + \frac{1}{R_s/R_b-1}\boldsymbol{r}_{b/s}$ \\

    $\displaystyle \boldsymbol{\rho}_p = \boldsymbol{r} - \boldsymbol{p}$ & 
    $\displaystyle \boldsymbol{\rho}_u = \boldsymbol{r} - \boldsymbol{u}$ \\

    \hline
\end{tabular}
\end{table}
\begin{table}[hbt!]
\caption{\label{tab:sem_eclipse} Sun-Earth-Moon Eclipse Geometry}
    \centering
    \begin{tabular}{lcccccc}
        \hline
        System    & $\alpha_p$ [deg]& $p_s$ [km]& $p_b$ [km] & $\alpha_u$ [deg]& $u_s$ [km]& $u_b$ [km]\\
        \hline
        Sun-Earth  & 0.2689 &  1.4826093$\times10^8$  & 1.3592474$\times10^6$
                   & 0.2640 &  1.5100457$\times10^8$  & 1.3844010$\times10^6$ \\
        Sun-Moon   & 0.2671 &  1.4925984$\times10^8$  & 3.7275268$\times10^5$
                   & 0.2657 &  1.5000721$\times10^8$  & 3.7461913$\times10^5$ \\
        \hline
    \end{tabular}
\end{table}
The antumbra is the umbra's dual cone, sharing its vertex and angle, but opening away from the secondary occluding body. The nominal umbral vertex for the Sun-Earth system lies far beyond the Earth-Moon system, while that of the Sun-Moon system lies roughly on the circle containing the rotating Earth-Moon triangular equilibrium points. While triangular libration point trajectories and many families of sidereal inertial resonance orbits span this region and could possibly encounter this particular lighting condition, it is rare, and thus the antumbra is omitted from the work developed in this paper. The remainder of this section discusses a simple eclipse detection scheme based on spacecraft telemetry, and the utility of an implicit measurement function based on second-order cone boundaries.

\subsection{Onboard Eclipse Detection}
Eclipse detection can be performed using several classes of onboard sensors commonly available on spacecraft. Temperature sensors such as thermistors and thermocouples can detect eclipse events through thermal transients; however, the thermal response of the spacecraft bus is strongly dependent on material properties, geometry, and active thermal control systems, making it difficult for a simple engineering model to perform accurately for navigation purposes. Sun sensors, including coarse and fine photodiode-based instruments, provide direct measurements of incident light but depend on spacecraft geometry and field-of-view constraints, which may limit their availability during certain attitudes. In this work, eclipse detection is based on monitoring electrical power generation from solar arrays, which provides a direct and continuous proxy for incident solar flux while remaining largely independent of spacecraft geometry under the assumption that single-axis solar array drive assemblies (SADA) continually maximize the available power generated given a bus attitude. Available power telemetry is standard across all classes of spacecraft because of its necessary role in operations and fault diagnosis, and its response to eclipse transitions is observable and straightforward to model.

The most basic model of generated solar power $\mathcal{P}(t)$ assumes an instantaneous mapping between illumination and power, expressed as
\begin{equation}
    \mathcal{P}(t)=\mathcal{P}_{\max}\gamma(t) \label{eq:power_raw}
\end{equation}
where $\gamma(t)$ is the shadow function representing the fraction of incident solar flux available at the spacecraft based on lighting conditions. Not explicitly present in Eq. \ref{eq:power_raw} are both the solar array alignment, and the solar intensity factor dependent on distance relative to the Sun. These factors are assumed constant to produce a constant maximum power term, but could be readily included in onboard implementation as their dependents are all typically accessible telemetry points on a standard spacecraft bus. This simple model is frequently used in low-thrust trajectory optimization and eclipse analysis \cite{pontani2022optimal,zhang2022comparison,zubko2024simplified}.

The shadow function $\gamma(t)$ is defined as the fraction of visible solar disk as seen from the spacecraft. Under the assumption of spherical emitting and occulting bodies, it is computed using the apparent angular geometry between the Sun and the occulting body. This formulation is standard in eclipse and solar radiation pressure modeling, where the overlap is computed using circle–circle intersection geometry \cite{montenbruck2002satellite}. The angular separation between the Sun and occulting body centers, as observed from the spacecraft, is
\begin{equation}
    \delta = \cos^{-1}\left(\frac{(\boldsymbol{r}_s-\boldsymbol{r})^\text{T}(\boldsymbol{r}_b-\boldsymbol{r})}{||(\boldsymbol{r}_s-\boldsymbol{r})||\,||(\boldsymbol{r}_b-\boldsymbol{r})||}\right)
\end{equation}
and the apparent angular radii are
\begin{align}
    \beta_s=\sin^{-1}\frac{R_s}{||\boldsymbol{r}_s-\boldsymbol{r}||},\quad
    \beta_b=\sin^{-1}\frac{R_b}{||\boldsymbol{r}_b-\boldsymbol{r}||}
\end{align}
The shadow function is then given by
\begin{align}
    \gamma(t)&= 1 - \frac{A_{\text{overlap}}}{\pi\beta_s^2}\\
    A_{\text{overlap}} &= \beta_s^2\cos^{-1}\left(\frac{\delta^2+\beta_s^2-\beta_b^2}{2\delta\beta_s}\right) + 
    \beta_b^2\cos^{-1}\left(\frac{\delta^2+\beta_b^2-\beta_s^2}{2\delta\beta_b}\right)\\ &- 
    \frac{1}{2}\sqrt{(\beta_s+\beta_b-\delta)(\beta_s-\beta_b+\delta)(-\beta_s+\beta_b+\delta)(\beta_s+\beta_b+\delta)} \nonumber
\end{align}

Alternative shadow models exist that account for higher-fidelity effects such as atmospheric refraction, limb darkening, and non-spherical body geometry. For example, Li et al. \cite{li2019shadow} incorporate atmospheric effects into the shadow function, while Song and Kim \cite{song2016effect} analyze the impact of Earth’s oblateness on eclipse prediction accuracy. These effects can shift predicted entry and exit times by several seconds up to a minute depending on geometry and altitude. Zhang et al. \cite{zhang2022comparison} further demonstrate that the choice of shadow model can considerably impact precise orbit determination during eclipse phases in Earth orbit. In this work, the spherical-body shadow model is retained for simplicity and mathematical tractability. Modeling discrepancies arising from neglected effects such as oblateness or atmospheric refraction are instead captured by the event-time uncertainty $\sigma_t$, which is derived from these physical sources in Sec. \ref{sec:4}. As will be shown in Sec. \ref{sec:5}, this approximation remains sufficient for reducing state uncertainty within the navigation filter for trajectories in cislunar space.

Given a power telemetry signal $\mathcal{P}(t)$, eclipse transition times can be directly extracted. Penumbral contacts can be estimated by fitting the shadow function $\gamma(t)$ to the samples on the illumination ramp, while umbral contacts can be estimated directly from the samples that bracket the interval of zero generated power. The uncertainty of the contact times is derived in Sec. \ref{sec:4}, and the umbral contact timing is compared against flight data in Sec. \ref{sec:5}. Depending on trajectory geometry, three meaningful detection scenarios may occur: full penumbra-umbra-penumbra traversal (four measurements), penumbra-only traversal (two measurements), or no shadow intersection.

\subsection{The Zero Measurement}
Unlike prior work in two-body low-altitude orbits where eclipse timing can be predicted analytically or semi-analytically under two-body dynamics \cite{neta1998satellite,fitzgerald2021autonomous,dewalch2024enhancing}, eclipse events in cislunar space must be determined numerically due to the non-periodic and highly perturbed geometry of the Sun–Earth–Moon system. Rather than directly treating detection time as the measurement, this work formulates eclipse events as implicit geometric conditions, defined by the spacecraft’s location relative to the eclipse cones. Specifically, a geometric measurement is constructed such that it evaluates to zero at the moment an eclipse transition is detected, thus providing spatiotemporal information about the spacecraft's state with respect to celestial bodies. For the penumbral and umbral cones, the measurement functions are given by
\begin{gather}
    h_p(\boldsymbol{r},t) = \boldsymbol{\rho}_p^\text{T}\hat{\boldsymbol{r}}_{b/s} - ||\boldsymbol{\rho}_p||\cos\alpha_p \label{eq:pen_m}\\
    h_u(\boldsymbol{r},t) = -\boldsymbol{\rho}_u^\text{T}\hat{\boldsymbol{r}}_{b/s} - ||\boldsymbol{\rho}_u||\cos\alpha_u \label{eq:umb_m}
\end{gather}
with quantities defined in Table \ref{tab:eclipse_eqs}. These expressions represent second-order cone boundaries, and satisfy $h_p=0$ and $h_u=0$ at the respective entry and exit events.

The measurement sensitivities required for first-order estimation are obtained by linearizing Eqs. \ref{eq:pen_m} and \ref{eq:umb_m} with respect to the spacecraft state. The partial derivatives with respect to position are
\begin{gather}
    \frac{\partial h_p}{\partial\boldsymbol{r}}=\hat{\boldsymbol{r}}_{b/s}^\text{T} - \frac{\boldsymbol{\rho}_p^\text{T}}{||\boldsymbol{\rho}_p||}\cos\alpha_p\label{eq:dhdr_p}\\
    \frac{\partial h_u}{\partial\boldsymbol{r}}=-\hat{\boldsymbol{r}}_{b/s}^\text{T} - \frac{\boldsymbol{\rho}_u^\text{T}}{||\boldsymbol{\rho}_u||}\cos\alpha_u\label{eq:dhdr_u}
\end{gather}
while the partials with respect to velocity are zero. As will be shown in Sec. \ref{sec:4}, the time derivatives of the zero-measurement functions are required to characterize the measurement noise. These are given by
\begin{gather}
    \frac{\partial h_p}{\partial t}=\dot{\boldsymbol{\rho}}_p^\text{T}\hat{\boldsymbol{r}}_{b/s} + \boldsymbol{\rho}_p^\text{T}\dot{\hat{\boldsymbol{r}}}_{b/s} - \frac{\boldsymbol{\rho}_p^\text{T}\dot{\boldsymbol{\rho}}_p}{||\boldsymbol{\rho}_p||}\cos\alpha_p\label{eq:dhdt_p}\\
    \dot{\boldsymbol{\rho}}_p = \boldsymbol{v} - \boldsymbol{v}_s - \frac{1}{1+R_b/R_s}\boldsymbol{v}_{b/s}\\
    \frac{\partial h_u}{\partial t}=-\left(\dot{\boldsymbol{\rho}}_u^\text{T}\hat{\boldsymbol{r}}_{b/s} + \boldsymbol{\rho}_u^\text{T}\dot{\hat{\boldsymbol{r}}}_{b/s}\right) - \frac{\boldsymbol{\rho}_u^\text{T}\dot{\boldsymbol{\rho}}_u}{||\boldsymbol{\rho}_u||}\cos\alpha_u\label{eq:dhdt_u}\\
    \dot{\boldsymbol{\rho}}_u = \boldsymbol{v} - \boldsymbol{v}_b - \frac{1}{R_s/R_b-1}\boldsymbol{v}_{b/s}
\end{gather}
where it is assumed that the occluding body's radial velocity relative to the Sun is much smaller than the spacecraft's velocity as it flies through the eclipse around event times. This greatly simplifies the measurement time derivative, as
\begin{equation}
    \dot{\alpha}_p\approx0, \quad \dot{\alpha}_u\approx0
\end{equation}
The time derivative of the eclipsing axis is given by
\begin{equation}
    \dot{\hat{\boldsymbol{r}}}_{b/s} = \frac{1}{{\boldsymbol{r}_{b/s}}^\text{T}\boldsymbol{r}_{b/s}}\left(||\boldsymbol{r}_{b/s}||\cdot\boldsymbol{v}_{b/s} - \frac{\boldsymbol{r}_{b/s}^\text{T}\boldsymbol{v}_{b/s}}{||\boldsymbol{r}_{b/s}||}
    \cdot\boldsymbol{r}_{b/s}\right),\quad \boldsymbol{v}_{b/s}=\boldsymbol{v}_b - \boldsymbol{v}_s
\end{equation}

To ensure physical consistency and reject spurious detections (e.g., due to sensor error or solar array misalignment), candidate detection times can be validated by evaluating the measurement residual using the propagated state and a planetary ephemeris model. Only events satisfying the appropriate cone-side condition are accepted. This implicit “zero-measurement” formulation provides a straightforward, differentiable representation of eclipse transitions, enabling direct integration into any batch or sequential filtering framework.

\section{Filter Integration and Event-Time Uncertainty} \label{sec:4}
The measurement model developed above is positioned to supplement any batch or sequential filter as an event-based update tracking an arbitrary spacecraft trajectory in the Sun-Earth-Moon system. Here, it is integrated for demonstration into a UKF to estimate a spacecraft's inertial state $\boldsymbol{x}=[\boldsymbol{r},\:\boldsymbol{v}]^\text{T}$ in multi-body cislunar orbits \cite{wan2000unscented}. While an extended Kalman filter is applicable for the $L_1$ halo example used below, the first-order covariance mapping breaks down during the perilune traverse in CAPSTONE's NRHO for the uncertainty sizes examined in this paper, thus a UKF at a minimum must be employed.

The onboard eclipse detection event is formulated as the implicit zero-measurement as seen in Sec. \ref{sec:3}, where the geometric measurement function $h(\boldsymbol{r}, t)$ is expected to evaluate to zero at the moment of penumbral or umbral transition. The measurement $z_k= h(\boldsymbol{r}(t_{k}), t_{k})+\nu_k$ is defined as the scalar value of the geometric cone boundary, where $\nu_k$ is assumed to be a zero-mean, Gaussian-distributed white noise process, $\nu_k \sim \mathcal{N}(0, R_k)$. Explicit measurement uncertainty of detection time $\sigma_t$ arises from a combination of available data rate, power signal noise, data-processing error, and unmodeled eclipsing effects like non-spherical occluding bodies and atmospheric refraction. This timing error is mapped into the geometric domain via the time-sensitivity of the zero-measurement function in Eqs. \ref{eq:dhdt_p} and \ref{eq:dhdt_u}. Thus, the event-specific measurement noise covariance $R_k$ is defined as
\begin{equation}
    \delta h \approx\frac{\partial h}{\partial t}\delta t\implies R_k=\mathbb{E}\left\{(\delta h)^2\right\}=\left(\left.\frac{\partial h}{\partial t}\right|_{t_k}\right)^2\sigma_t^2\label{eq:Rk}
\end{equation}
Finally, the measurement partials are
\begin{equation}
    \boldsymbol{H}_k=\left.\frac{\partial \boldsymbol{h}}{\partial\boldsymbol{x}}\right|_{t_k}=
    \left[\left.\frac{\partial h}{\partial\boldsymbol{r}}\right|_{t_k}\quad \boldsymbol{0}_{1\times3}\right]
\end{equation}
with position sensitivities found in Eqs. \ref{eq:dhdr_p} and \ref{eq:dhdr_u}. This formulation ensures that a filter appropriately weights the measurement based on the relative velocity and geometry of the specific eclipse crossing.


The event-time uncertainty $\sigma_t$ in Eq. \ref{eq:Rk} must account for the physical effects that the spherical shadow model and a sampled power signal do not capture. The effects considered here are the sampling of the power telemetry and the departure of the occulting body from the sphere of radius $R_b$ assumed in Table \ref{tab:eclipse_eqs}.

\subsection{Event-Time Error Model}\label{sec:sigt}
The geometric contact time $t_k$ satisfies $h(\boldsymbol{r}(t_k),t_k;R_b)=0$, where the dependence of the measurement function on the occulting body radius $R_b$ through Table \ref{tab:eclipse_eqs} is now written explicitly. If the effective radius of the occulting body differs from the modeled value by $\delta R_b$, the contact occurs at $t_k+\delta t$ instead. Linearizing $h=0$ about the modeled contact in the same way as Eq. \ref{eq:Rk} gives
\begin{equation}
    \frac{\partial h}{\partial t}\delta t+\frac{\partial h}{\partial R_b}\delta R_b=0\implies\delta t=-\left(\frac{\partial h}{\partial R_b}\middle/\frac{\partial h}{\partial t}\right)\delta R_b\label{eq:dt_dRb}
\end{equation}
The partial derivatives of Eqs. \ref{eq:pen_m} and \ref{eq:umb_m} with respect to $R_b$ follow from the relations in Table \ref{tab:eclipse_eqs}, using $\sin\alpha_p=(R_s+R_b)/||\boldsymbol{r}_{b/s}||$ and $\sin\alpha_u=(R_s-R_b)/||\boldsymbol{r}_{b/s}||$ from the same table. Evaluated on the cone boundaries, where $h_p=0$ and $h_u=0$, they are
\begin{gather}
    \frac{\partial h_p}{\partial R_b}=\frac{R_s+||\boldsymbol{\rho}_p||\tan\alpha_p}{||\boldsymbol{r}_{b/s}||}\label{eq:dhdRb_p}\\
    \frac{\partial h_u}{\partial R_b}=\frac{R_s-||\boldsymbol{\rho}_u||\tan\alpha_u}{||\boldsymbol{r}_{b/s}||}\label{eq:dhdRb_u}
\end{gather}
On the cone boundaries, the magnitude of the position partials in Eqs. \ref{eq:dhdr_p} and \ref{eq:dhdr_u} is exactly $\sin\alpha$. At the contacts examined in Sec. \ref{sec:5}, Eqs. \ref{eq:dhdRb_p} and \ref{eq:dhdRb_u} are within 0.23\% of $\sin\alpha$ for the $L_1$ halo and within 0.005\% for CAPSTONE. An error $\delta R_b$ in the occulting body radius is therefore equivalent to a position error of nearly the same size normal to the cone boundary.

Telemetry sampling adds a timing error that is independent of $\delta R_b$. Combining the sampling term of Sec. \ref{sec:samp} with Eq. \ref{eq:dt_dRb}, the event-time variance of contact $k$ and the corresponding measurement noise covariance from Eq. \ref{eq:Rk} are
\begin{gather}
    \sigma_{t,k}^2=\frac{\Delta t_s^2}{12}+\left(\frac{\partial h}{\partial R_b}\middle/\frac{\partial h}{\partial t}\right)^2_{t_k}\sigma_{R,k}^2\label{eq:sigt}\\
    R_k=\left(\left.\frac{\partial h}{\partial t}\right|_{t_k}\right)^2\frac{\Delta t_s^2}{12}+\left(\left.\frac{\partial h}{\partial R_b}\right|_{t_k}\right)^2\sigma_{R,k}^2\label{eq:Rk_budget}
\end{gather}
where $\Delta t_s$ is the telemetry sample interval and $\sigma_{R,k}$ is the standard deviation of the effective occulting body radius at contact $k$. For Earth, $\sigma_{R,k}^2=\sigma_{R,\text{atm}}^2+\sigma_{R,\text{obl}}^2$, the contributions of the atmosphere and of oblateness. For the Moon, $\sigma_{R,k}=\sigma_{R,\text{topo}}$, the contribution of limb topography. Note that the radius term in $R_k$ does not depend on the crossing rate $\partial h/\partial t$, while its contribution to $\sigma_t$ does. Each term is sized below for $\Delta t_s=15$ s, representative of the CAPSTONE power telemetry in Sec. \ref{sec:5}.

\subsection{Telemetry Sampling}\label{sec:samp}
A contact that is located to within one telemetry interval, with arbitrary phase relative to the sampling clock, carries a uniformly distributed timing error with standard deviation $\Delta t_s/\sqrt{12}$ \cite{widrow2008quantization}. At $\Delta t_s=15$ s for CAPSTONE, this is 4.33 s, independent of the detection scheme. A delay in power system response path from photon impact to readable data point is assumed to be a constant offset that can be calibrated on the ground. For the purpose of this work, it is assumed that this delay time falls well below the telemetry interval, and is thus negligible in that regard.

\subsection{Effective Occulting Body Radius}

\subsubsection{Earth's Atmosphere}
Atmospheric refraction and extinction of grazing sunlight make Earth's shadow larger than the geometric shadow of the solid Earth \cite{li2019shadow}. Herald and Sinnott \cite{herald2014analysis} reduced 22,539 timing records of lunar craters and limb features entering and leaving the umbra at 94 lunar eclipses between 1842 and 2011 using an oblate Earth plus an eclipse-forming layer, and found a mean layer height of $86.9\pm0.2$ km that does not vary with lunar distance over 56 to 64 Earth radii, with the solar cycle, or with season. This mean layer is a one-signed offset of approximately 91 to 92 s at the halo umbral contacts of Sec. \ref{sec:5}, which zero-mean noise cannot represent. To account for this deterministically, the umbral radius is computed using $R_b=R_\oplus+h$ to absorb this offset while keeping the shadow model spherical. In the simulation of Sec. \ref{sec:5}, the true contacts are generated with the same spherical model used in the filter, so only the eclipse-to-eclipse scatter of the layer height enters $\sigma_t$. From the catalog accompanying \cite{herald2014analysis}, the standard deviation of the per-eclipse mean layer height, corrected for the finite number of timings per eclipse, is 4.5 to 6.1 km for the eclipses with at least 100 timings, depending on whether the sparsely observed eclipses of 1943 to 1974 are excluded, so $\sigma_{R,\text{atm}}=6$ km is used. No layer term is applied to penumbral contacts, since an enlargement of the outer edge of the penumbra is not observed \cite{herald2014analysis}. Atmospheric refraction still reshapes the illumination profile near Earth's limb within the penumbra, but this effect is not quantified by the lunar eclipse data and is not included here.

\subsubsection{Earth's Oblateness}
Earth's oblateness changes the shape of its silhouette as seen from the Sun. For a spheroid with equatorial radius $a$, flattening $f$, and polar radius $b=a(1-f)$, the projection along $\hat{\boldsymbol{r}}_{b/s}$ is an ellipse with semimajor axis $a$ and semiminor axis $b'$, where
\begin{equation}
    b'^2=a^2\sin^2\delta_\odot+b^2\cos^2\delta_\odot
\end{equation}
and $\delta_\odot$ is the declination of the Sun in Earth's equatorial frame. To first order in $f$, $b'\approx a(1-f\cos^2\delta_\odot)$, and the radius of the silhouette at position angle $\theta$, measured about $\hat{\boldsymbol{r}}_{b/s}$ from the equatorial direction, is
\begin{equation}
    R(\theta)\approx a\left(1-f\cos^2\delta_\odot\sin^2\theta\right)\label{eq:obl}
\end{equation}
Herald and Sinnott \cite{herald2014analysis} use the same form for Earth's umbra, with an effective flattening that varies from about 1/298 near equinox to about 1/355 near solstice. The equatorial radius used in this work is exact at $\theta=0^\circ$ and $180^\circ$. Since $\theta$ follows from the ephemeris at each contact, Eq. \ref{eq:obl} could be applied directly. Here, $\theta$ is instead treated as unknown and uniformly distributed to account for all configurations. Then $\sin^2\theta$ has mean 1/2 and variance 1/8, so the silhouette radius has a mean offset of $af\cos^2\delta_\odot/2$ from the equatorial radius. This can be removed using the mean radius, and a standard deviation of
\begin{equation}
    \sigma_{R,\text{obl}}=\frac{af\cos^2\delta_\odot}{2\sqrt{2}}
\end{equation}
With the GRS80 values $a=6378.137$ km and $1/f=298.257222101$ \cite{petit2010iers}, $af=21.385$ km and $\sigma_{R,\text{obl}}=6.69$ km at the halo eclipse of Sec. \ref{sec:5}, where $\delta_\odot=-19.8^\circ$. The halo contacts occur within $7^\circ$ of Earth's equatorial limb, where the equatorial radius is accurate to within 0.25 km, so this value is conservative for the halo example. For comparison, Song and Kim \cite{song2016effect} found spherical and spheroidal Earth shadow predictions for a lunar-orbiting spacecraft to differ by about half a minute on average.

\subsubsection{Lunar Topography}
The Moon is modeled as a 1737.4 km sphere. Features on the lunar limb depart from a sphere by about $\pm6$ km \cite{espenaklimb}, within the global relief of $+10.78$ to $-9.12$ km measured by the Lunar Orbiter Laser Altimeter (LOLA) \cite{smith2010initial}. A published root-mean-square value of the limb profile height was not found, so $\sigma_{R,\text{topo}}=2$ km is adopted as a $1\sigma$ value within the $\pm6$ km limb departure. The $\pm6$ km bound corresponds to 9.6 and 10.0 s at the CAPSTONE umbral entry and exit contacts.

\subsection{Event-Time Uncertainty Budget}
\begin{table}[hbt!]
\caption{\label{tab:sigt_budget} Event-Time Uncertainty Budget at a 15 s Telemetry Interval}
    \centering
    \footnotesize
    \begin{tabular}{lccccccc}
        \hline
        Contact &  {$|\partial h/\partial t|$ [m/s]} &  {$\partial h/\partial R_b$ [$10^{-3}$]} &  {Sampling [s]} & Atmosphere [s] &  {Shape [s]} &  {RSS [s]} &  {$\sigma_t$ [s]} \\
        \hline
        Halo penumbral entry & 4.610 & 4.781 &  {4.33} &  {--}   &  {6.94} &  {8.18}  &  {8} \\
         {Halo umbral entry}    &  {4.485} &  {4.694} &  {4.33} &  {6.28} &  {7.01} &  {10.36} &  {10} \\
         {Halo umbral exit}     &  {4.420} &  {4.694} &  {4.33} &  {6.37} &  {7.11} &  {10.48} &  {10} \\
         {Halo penumbral exit}  &  {4.478} &  {4.781} &  {4.33} &  {--}   &  {7.15} &  {8.36}  &  {8} \\
         {CAPSTONE umbral entry}&  {2.946} &  {4.720} &  {4.33} &  {--}   &  {3.20} &  {5.39}  &  {5} \\
         {CAPSTONE umbral exit} &  {2.828} &  {4.720} &  {4.33} &  {--}   &  {3.34} &  {5.47}  &  {5} \\
        \hline
    \end{tabular}
\end{table}
The assembled budget is given in Table \ref{tab:sigt_budget}, where each term follows from Eq. \ref{eq:sigt} and the shape term is $\sigma_{R,\text{obl}}$ for the halo and $\sigma_{R,\text{topo}}$ for CAPSTONE. Rounded to whole seconds, $\sigma_t$ is 8 s at the halo penumbral contacts, 10 s at the halo umbral contacts, and 5 s at the CAPSTONE umbral contacts. The CAPSTONE value is consistent with the flight umbral contact residuals of $+0.37$ s and $+6.04$ s found in Sec. \ref{sec:5}. The halo crosses the cone boundaries faster than CAPSTONE, with $|\partial h/\partial t|$ of 4.4 to 4.6 m/s against 2.8 to 2.9 m/s. Consequently, the larger halo event-time uncertainty comes from the larger radius error of Earth, about 9 km at the umbral contacts against 2 km for the Moon.

Within one complete eclipse transition, the occulting body terms are common to all contacts, and not independent. However, the filter employed in Sec. \ref{sec:5} treats each $R_k$ as independent, which overstates the information gained from multiple contacts of the same eclipse by choosing to ignore the correlation. This could be addressed in future work or implementation by estimating a per-eclipse timing bias, using a full measurement covariance with off-diagonal terms, or simply inflating $R_k$.

\section{Numerical Simulation}\label{sec:5}

\subsection{$L_1$ Halo Earth Eclipse}

\begin{figure}[htpb!]
    \centering
    \includegraphics[width=0.9\linewidth]{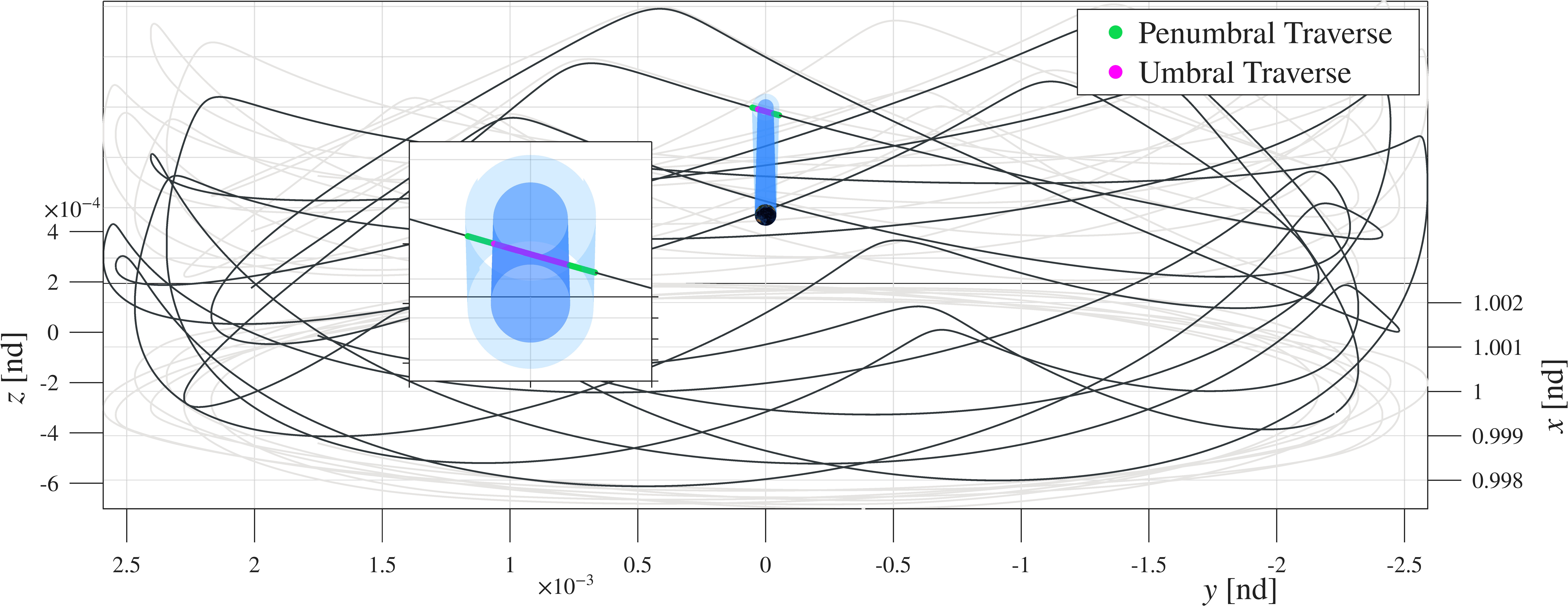}
    \caption{$L_1$ halo with Earth eclipse events in the Sun-Earth rotating-pulsating frame.}
    \label{fig:l1_data1}
\end{figure}

\begin{figure}[htpb!]
    \centering
    \includegraphics[width=0.5\linewidth]{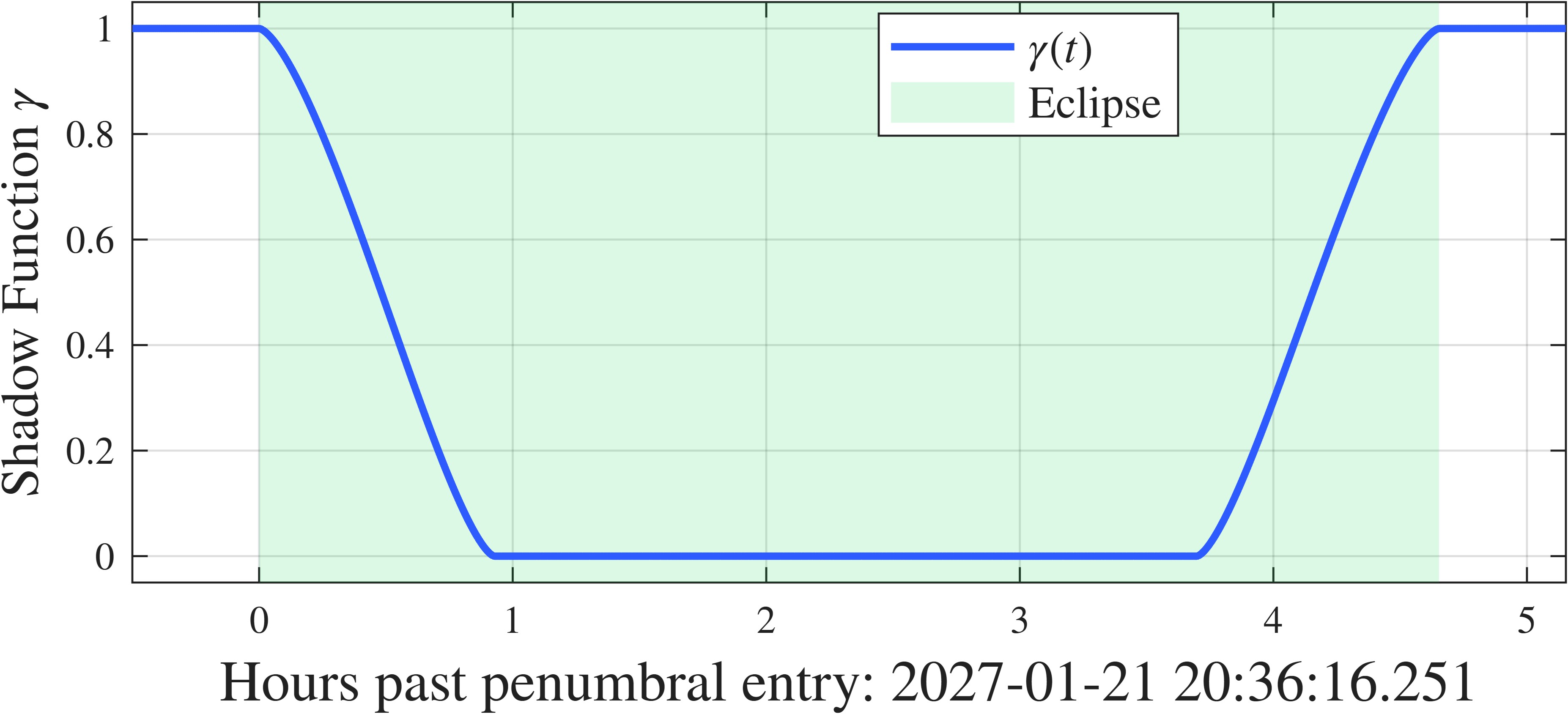}
    \caption{Spherical eclipse shadow function for $L_1$ halo Earth eclipse.}
    \label{fig:l1_data2}
\end{figure}

An $L_1$ southern Earth-Moon halo orbit with Jacobi constant 3.03617 and nondimensional period 2.72918 was computed in the CR3BP using the mass parameter $\bar{\mu}=0.01215$. It was then transitioned to the Sun-Earth-Moon point-mass ephemeris model described in Sec. \ref{sec:2} with $n=7$ patch points per revolution, and $m=21$ revolutions. The initial epoch was adjusted until a well defined Earth eclipse transition was located. The final converged trajectory, shown in Fig. \ref{fig:l1_data1}, has an initial epoch of 2026-09-06 20:23:43.927 with a total time of flight of 136.63 days.
There is a clear Earth occluded penumbral and umbral traversal seen in the Sun-Earth rotating frame that gives the opportunity for four event-based measurement updates to a sequential navigation filter. The spherical shadow function $\gamma(t)$ along the trajectory is also shown in Fig. \ref{fig:l1_data2}.
The true penumbral entry occurs at 2027-01-21 20:36:16.251 UTC, and umbral entry, umbral exit, and penumbral exit follow at 0.9276, 3.6989, and 4.6534 hours past penumbral entry, respectively. The true time spent in total shadow is 2.7713 hours, while the time spent in at least partial shadow is 4.6534 hours.

\begin{figure}[htbp!]
     \centering
     \begin{subfigure}{0.48\textwidth}
         \centering
         \includegraphics[width=1\textwidth]{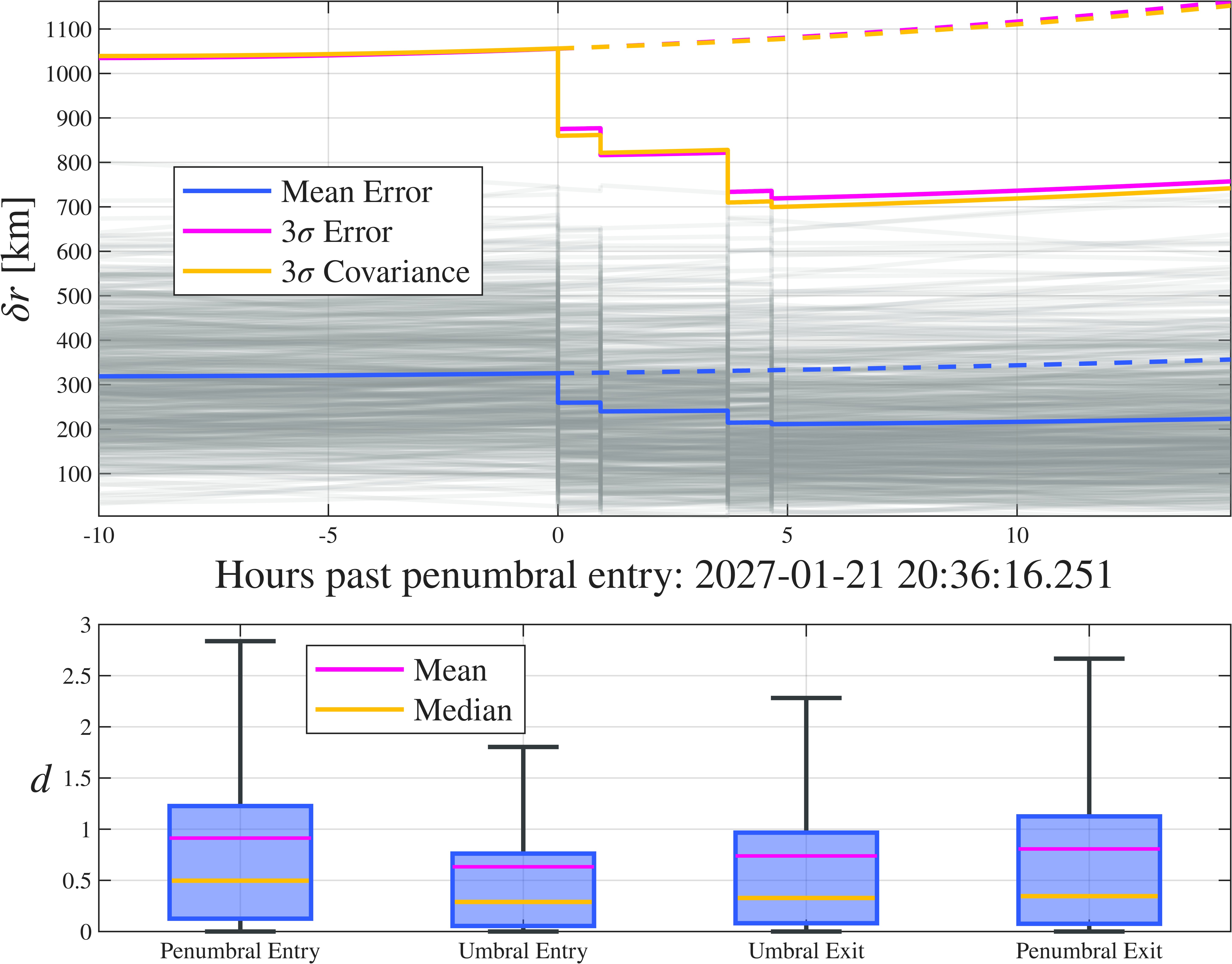}
         \caption{$(\sigma_t=8,\:10\text{ s},\,\sigma_r=346\text{ km})$ Monte Carlo Position Error and Measurement Mahalanobis Distance}
         \label{fig:l1_results_a}
     \end{subfigure}\hfill
     \begin{subfigure}{0.48\textwidth}
         \centering
         \includegraphics[width=1\textwidth]{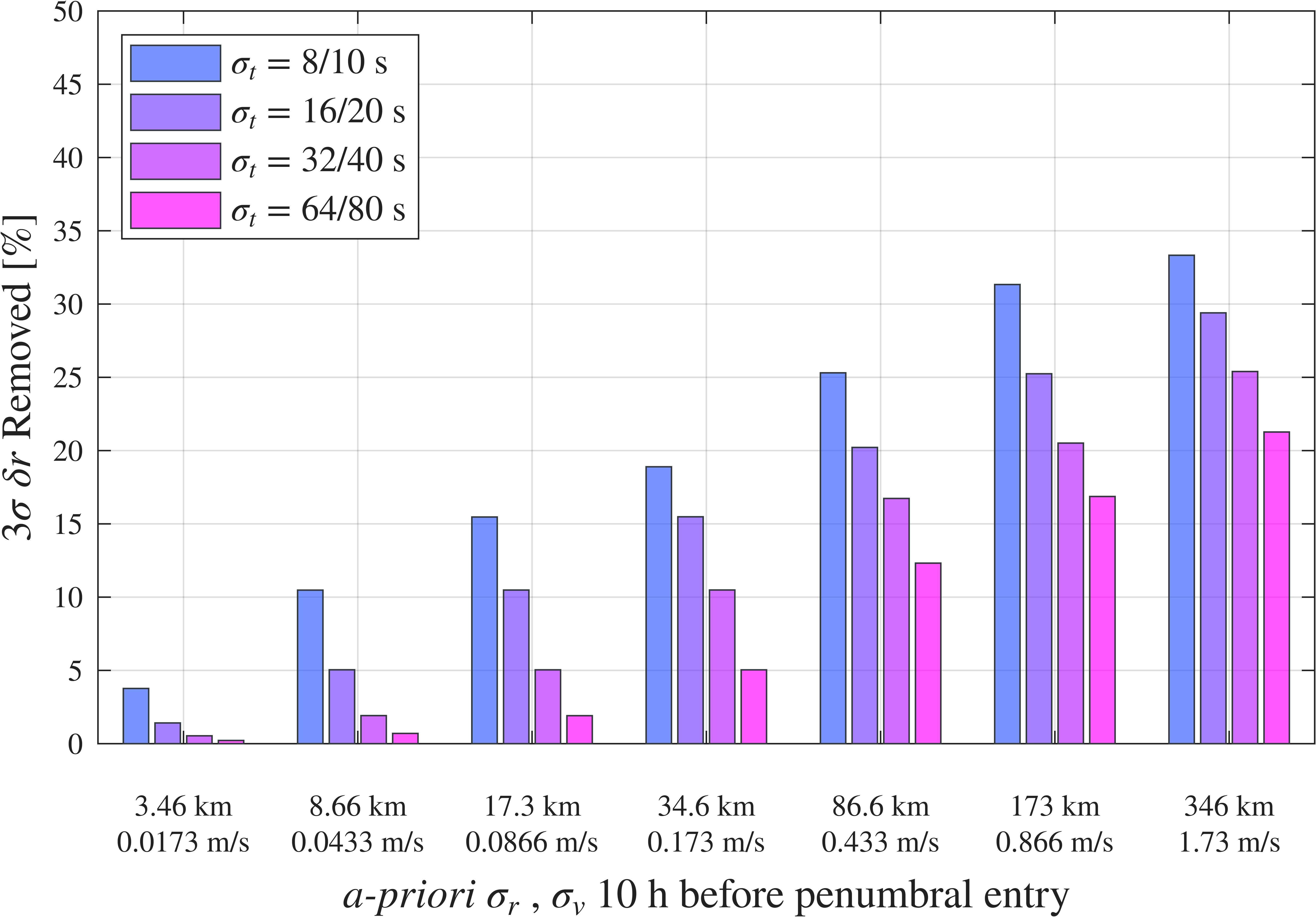}
         \caption{Average Position Uncertainty Reduction}
         \label{fig:l1_results_b}
     \end{subfigure}
     \caption{Uncertainty reduction study results for $L_1$ halo penumbral/umbral transition measurement processing during simulated Earth eclipse. Dashed lines show open-loop growth.}
     \label{fig:l1_results}
\end{figure}

To verify the utility of the eclipse-based measurements in a navigation filter, a 500 trial Monte Carlo simulation was run with a UKF. In each trial, the measured contact times are drawn about the true contact times with the per-contact event-time uncertainty of Table \ref{tab:sigt_budget}, $\sigma_t=8$, 10, 10, and 8 s, and the same $\sigma_t$ is declared in $R_k$ making the filter consistent by construction. The UKF is initialized 10 hours prior to true penumbral entry with an isotropically partitioned covariance matrix constructed with $\sigma=200\:\text{km},\:1\:\text{m/s}$ for each respective diagonal entry such that $\sigma_r=346\:\text{km}$ and $\sigma_v=1.73\:\text{m/s}$. Process noise is injected at the acceleration level with isotropic power spectral density $\boldsymbol{Q}=\texttt{diag}(0,0,0,q_a,q_a,q_a)$ where $q_a=1\times10^{-14}$ km$^2$/s$^3$. The truth trajectories are propagated in the same point-mass ephemeris model without random accelerations, so $q_a$ serves only as filter tuning. All four eclipse transition measurements are processed. Simulation results appear in Fig. \ref{fig:l1_results_a}.
\noindent In addition to the statistics for position magnitude error $\delta r$, Fig. \ref{fig:l1_results_a} also displays the box-and-whisker statistics for the $\chi^2$-distributed Mahalanobis distance $d$, the normalized innovation which quantifies information gain health \cite{mahalanobis2018generalized}. For the 1-dimensional zero-measurements of Eqs. \ref{eq:pen_m} and \ref{eq:umb_m}, this simplifies to
\begin{equation}
    d=\bar{h}_k^2 /S_k
\end{equation}
where $\bar{h}_k$ is the mean zero-measurement and $S_k$ is the measurement covariance, both computed from the unscented transform \cite{wan2000unscented}.

At the last contact, the $3\sigma$ position error drops from 1078.59 km for the open-loop baseline (dashed lines) to 718.81 km, a 33.4\% reduction, and the $3\sigma$ velocity error drops from 5.714 m/s to 4.944 m/s. The mean estimated $3\sigma$ position covariance at the last contact is 699.42 km, a ratio of sample error to covariance of 1.03, and the mean Mahalanobis distances at the four contacts are 0.91, 0.63, 0.74, and 0.81. This indicates statistical consistency and confirms the filter remains stable, avoiding overconfidence under realistic initialization and process noise injection. Ten hours after the last contact, the $3\sigma$ position error is 757.29 km against 1162.38 km open loop. The largest single reduction occurs at the initial penumbral entry update, which collapses the state uncertainty along the direction normal to the local cone boundary surface. Because the spacecraft's trajectory and the orientation of the shadow cones change very little over the 4.65-hour traversal, the cone normals of the four contacts mapped to the filter epoch lie within 2 to $11^\circ$ of one another. The subsequent transitions therefore provide constraints that are highly correlated with the first, and each yields a smaller reduction than penumbral entry.

An additional study was done to analyze the information gained from processing the four contacts given differing scales of initial uncertainty and event-time uncertainty. A 500 trial Monte Carlo simulation was run for each of seven initial uncertainty distributions, with diagonal $\sigma$ entries scaled from the 200 km, 1 m/s case by factors of 1, 0.5, 0.25, 0.1, 0.05, 0.025, and 0.01, and for the $\sigma_t$ budget of Table \ref{tab:sigt_budget} scaled by 1, 2, 4, and 8. The injected and declared $\sigma_t$ are scaled together, to maintain filter consistency. All cases use the same simulation start time and process noise as above with uncertainty reduction evaluated at the true penumbral exit epoch. The percentage of the open-loop $3\sigma$ position error removed is plotted in Fig. \ref{fig:l1_results_b}, where the $x$-axis labels are the root-sum-square $\sigma_r$ and $\sigma_v$ of each initial isotropic covariance matrix.

At the largest initial uncertainty (346 km, 1.73 m/s), the four contacts remove 33.3\% of the $3\sigma$ position error at the budget $\sigma_t$ and 21.3\% at eight times the budget. The reduction decreases with the initial uncertainty; at 3.46 km and 0.0173 m/s it is 3.8\% at the budget $\sigma_t$ and 0.2\% at eight times the budget. Filter consistency over the full state is checked with the normalized estimation error squared (NEES), $\epsilon=(\hat{\boldsymbol{x}}-\boldsymbol{x})^\text{T}\boldsymbol{P}^{-1}(\hat{\boldsymbol{x}}-\boldsymbol{x})$, computed for each trial at the evaluation epoch from the estimation error and the $6\times6$ filter covariance \cite{crassidis2004optimal}. For a consistent filter, $\epsilon$ follows a $\chi^2$ distribution with six degrees of freedom, with a median of 5.35 and a 95\% value of 12.59. The median NEES lies between 2.8 and 5.7 across all cases, with 0 to 6\% of trials above 12.59. A reduction of at least 25\% at the budget $\sigma_t$ requires an initial position uncertainty of 86.6 km or more. Fig. \ref{fig:l1_results_b} demonstrates the utility of processing the eclipse transition measurements when position uncertainty is on the order of $100$ km for this $L_1$ halo example.

\subsection{CAPSTONE Lunar Eclipse}
\begin{figure}[htbp!]
     \centering
     \begin{subfigure}{0.4\textwidth}
         \centering
         \includegraphics[width=1\textwidth]{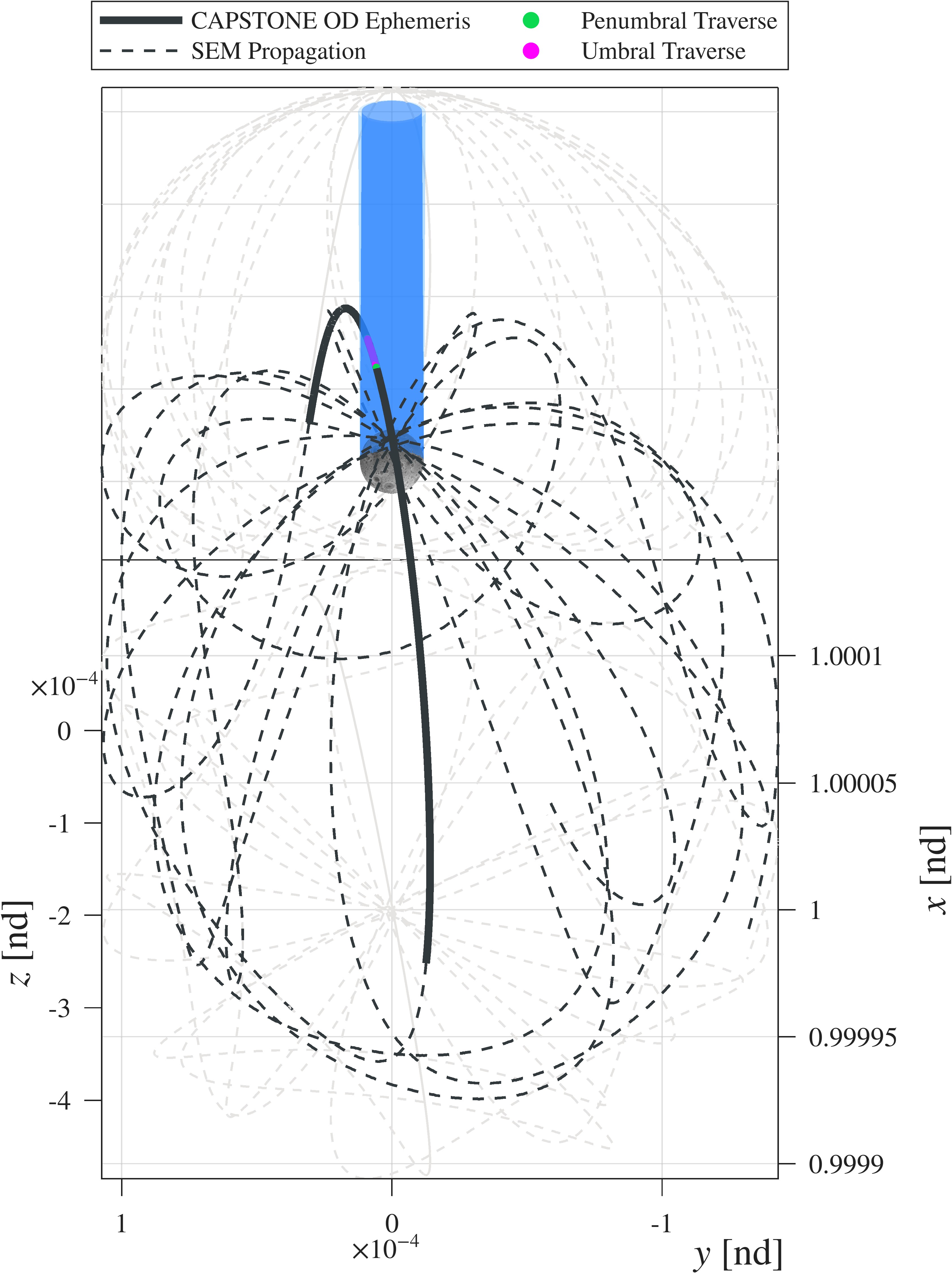}
         \caption{Sun-Moon Rotating-Pulsating Frame}
         \label{fig:capstone_data_a}
     \end{subfigure}\hfill
     \begin{subfigure}{0.52\textwidth}
         \centering
         \includegraphics[width=1\textwidth]{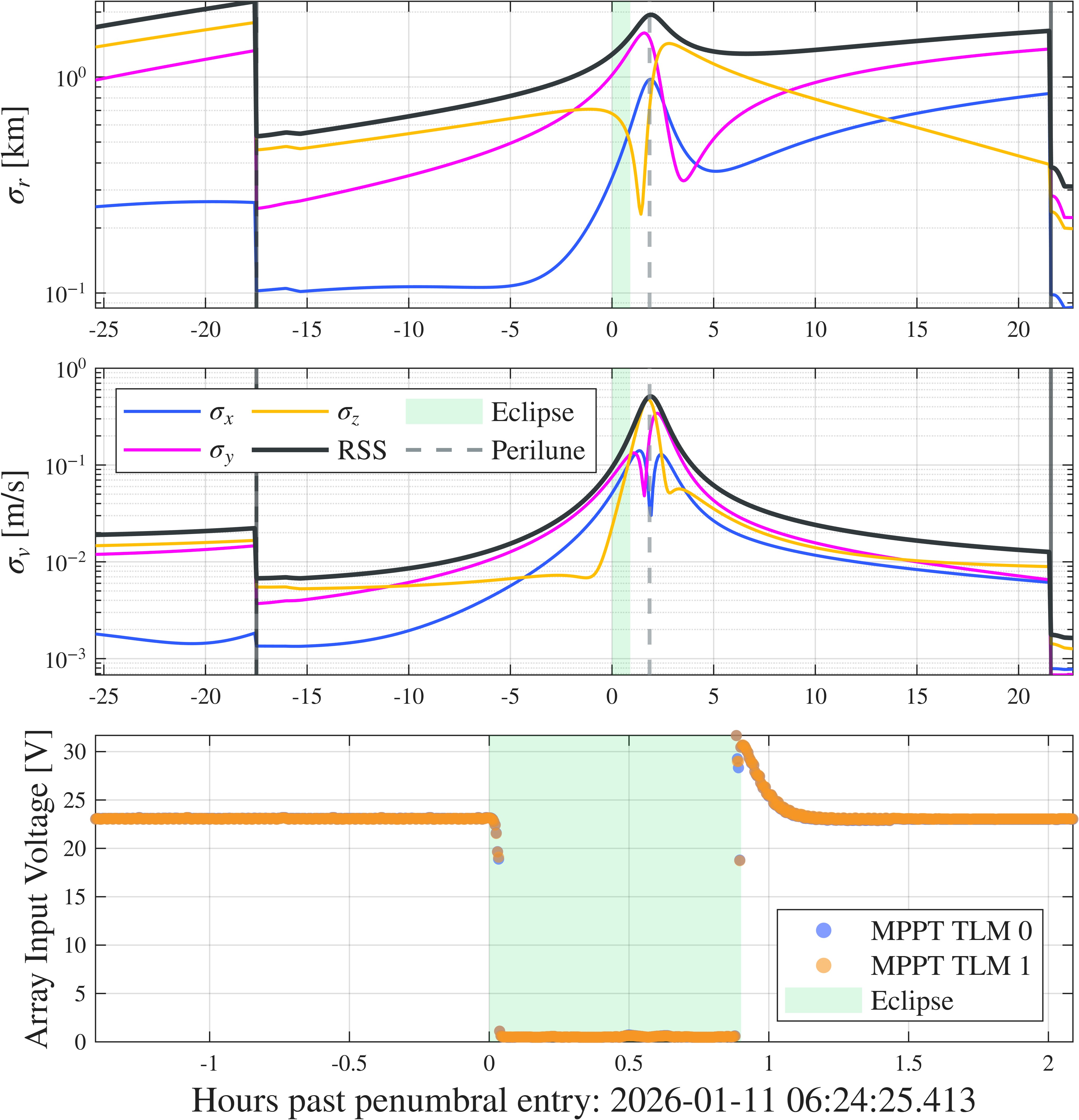}
         \caption{CAPSTONE OD Solution and Power Telemetry}
         \label{fig:capstone_data_b}
     \end{subfigure}
     \caption{CAPSTONE spacecraft ephemeris solution and power telemetry surrounding the January 2026 lunar eclipse.}
     \label{fig:capstone_data}
\end{figure}

Real telemetry and orbit determination (OD) solutions pulled from the CAPSTONE spacecraft are now used to demonstrate the utility of processing eclipse observables. CAPSTONE passed through the Moon's shadow on 2026-01-11 while inbound to perilune in its NRHO. The navigation products delivered for this event are the OD ephemeris in the EME2000 frame relative to Earth on a 5 minute grid, the mapped $6\times6$ OD covariance over the same span, and power telemetry from the spacecraft's two maximum power point tracking (MPPT) units at a 14.9 s interval, spanning 1.41 hours before to 2.09 hours after penumbral entry. Computed from the delivered ephemeris in the Sun-Earth-Moon dynamic model, penumbral entry occurs at 06:24:25.413 UTC, umbral entry at 06:26:18.322 UTC, umbral exit at 07:17:12.658 UTC, and penumbral exit at 07:18:28.566 UTC, followed by perilune at 08:15:14 UTC at an altitude of 1589 km. The spacecraft is 7767 km from the Moon at penumbral entry and 5154 km at penumbral exit, so the penumbral annulus is crossed in only 113 s at entry and 76 s at exit. The trajectory is shown in the Sun-Moon rotating-pulsating frame in Fig. \ref{fig:capstone_data_a}.

The telemetry time tags are in GPS time, and converting them to UTC removes an 18 s offset. With this conversion, the umbral contacts timed from the power telemetry agree with the geometric contacts computed from the delivered ephemeris to within a single sample interval. Refitting each 5 minute interval of the delivered ephemeris with a ballistic arc in the point mass ephemeris dynamic model of Eq. \ref{eq:ephem} presents a maximum velocity magnitude mismatch of 22.1 mm/s, occurring at perilune and falling off smoothly on either side. This is consistent with higher order lunar gravity field effects that are modeled in the delivered ephemeris but not in the point mass model. The Monte Carlo truth trajectory is therefore generated by propagating a single delivered state, which reproduces the delivered ephemeris to within 0.31 km at penumbral entry and 0.45 km at penumbral exit.

The delivered covariance in Fig. \ref{fig:capstone_data_b} shows two tracking updates in the window, at 2026-01-10 12:54:51 UTC, where $\sigma_r$ drops from 2.242 km to 0.533 km, and at 2026-01-12 03:59:51 UTC, 39 hours later. The eclipse occurs 17.49 hours after the first update. Between updates, $\sigma_r$ grows from 0.533 km to 1.274 km at penumbral entry, while $\sigma_v$ grows from 6.8 mm/s to 94 mm/s at penumbral entry and 510 mm/s at perilune. The array input voltage of both MPPT units, also shown in Fig. \ref{fig:capstone_data_b}, sits near 23 V in sunlight, collapses to approximately 0.5 V inside the umbra, and recovers with a 31 V overshoot after umbral exit. The transient behavior is produced as the solar arrays become temporarily more efficient immediately following a cool-down during the umbral pass.

\begin{figure}[htbp!]
     \centering
     \begin{subfigure}{0.48\textwidth}
         \centering
         \includegraphics[width=1\textwidth]{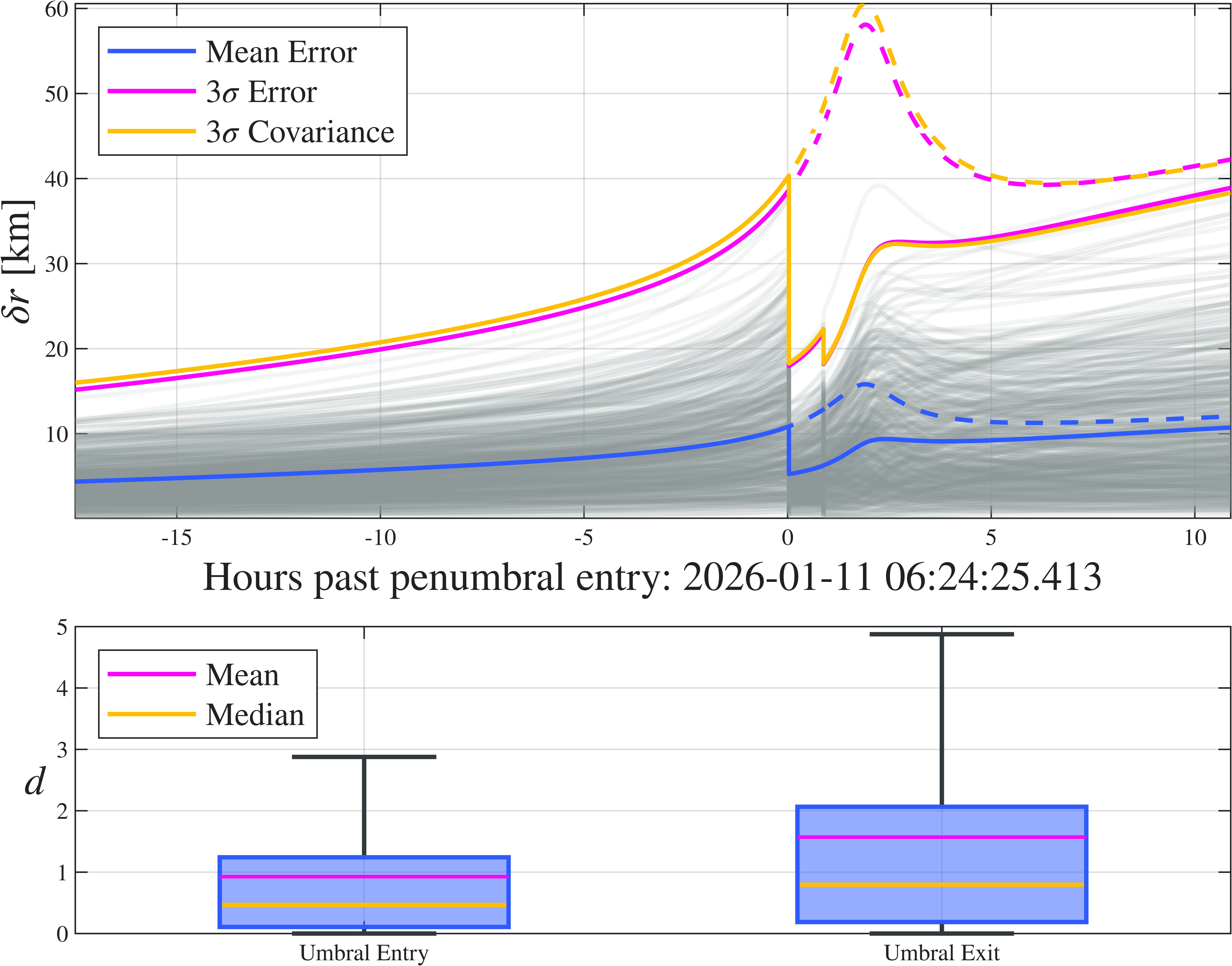}
         \caption{$(\sigma_t=5\text{ s},\,\sigma_r=\edit{5.33}\text{ km})$ Monte Carlo Position Error and Measurement Mahalanobis Distance}
         \label{fig:capstone_results_a}
     \end{subfigure}\hfill
     \begin{subfigure}{0.48\textwidth}
         \centering
         \includegraphics[width=1\textwidth]{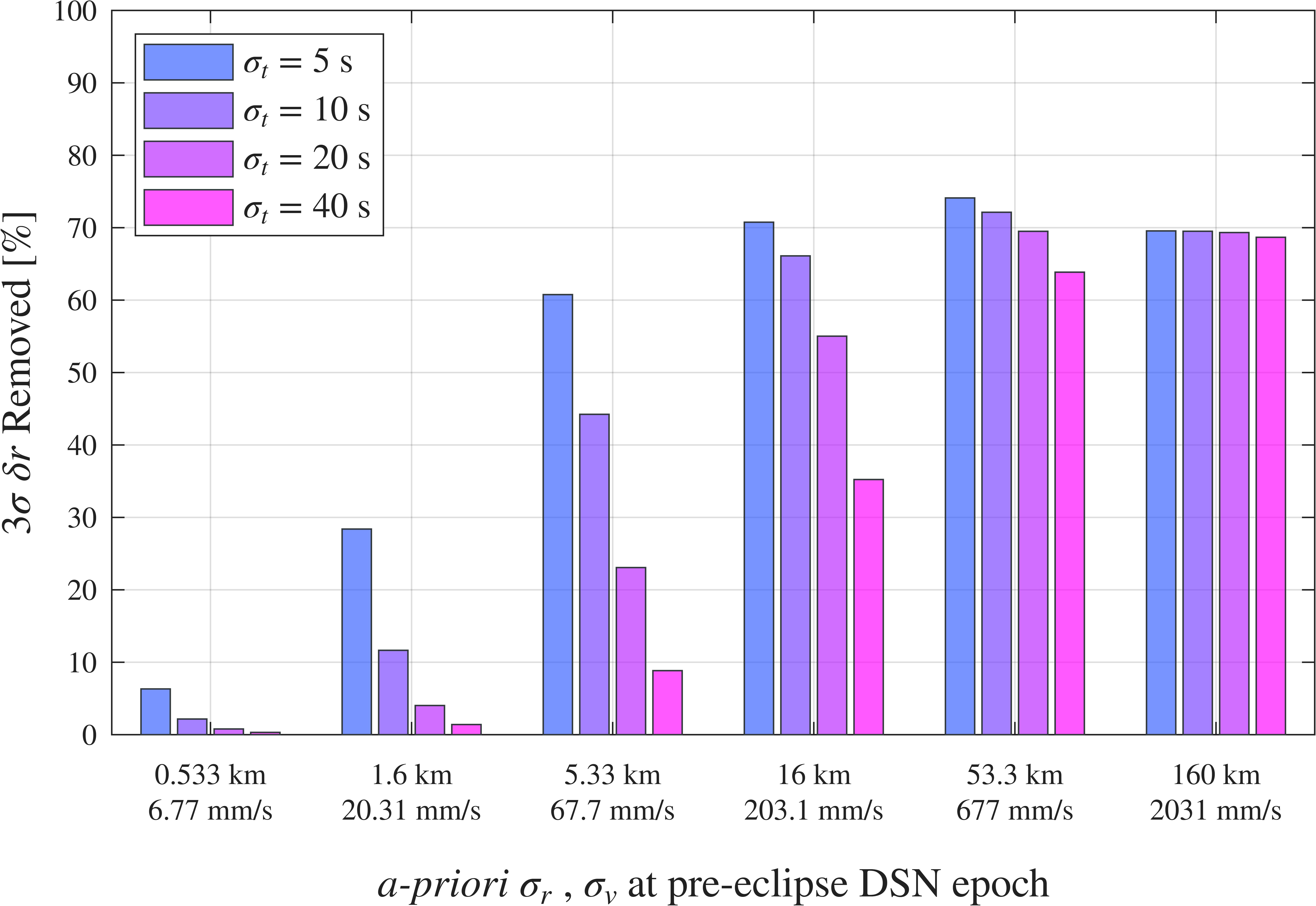}
         \caption{Average Position Uncertainty Reduction}
     \end{subfigure}
     \caption{Uncertainty reduction study results for CAPSTONE spacecraft umbral transition measurement processing during the January 2026 lunar eclipse.}
     \label{fig:capstone_results_b}
     \label{fig:capstone_results}
\end{figure}

Looking at Fig. \ref{fig:capstone_data_b}, there is no geometric information to be gained from processing both penumbral and umbral contacts because the time between the two on each side of the eclipse is simply to small, a product of passing so close to the eclipsing body. Therefore, only umbral measurements are processed in this example. The umbral contacts are timed directly from the telemetry by the midpoint of the pair of samples that brackets the interval of zero generated power. The resulting errors relative to the geometric contacts from the propagated ephemeris model are $+0.37$ s at umbral entry and $+6.04$ s at umbral exit, a root-mean-square of 4.3 s that conveniently matches the $\Delta t_s/\sqrt{12}=4.33$ s sampling term of Table \ref{tab:sigt_budget}. Using the first and last zero-power sample instead of the midpoint gives $+7.79$ s and $-1.47$ s. The value $\sigma_t=5$ s from Table \ref{tab:sigt_budget} is used for measurement processing in the UKF.

The initial UKF epoch is the DSN tracking update at 2026-01-10 12:54:51 UTC, 17.49 hours before penumbral entry, where the delivered covariance is a posterior with $\sigma_r=0.533$ km and $\sigma_v=6.77$ mm/s. Process noise is set to zero, since propagating the epoch covariance without process noise reproduces the delivered $\sigma_r$ at penumbral entry to within 5\% (1.338 km against 1.274 km). To represent a degraded navigation solution, such as after a missed tracking pass, the full delivered $6\times6$ covariance at the epoch is scaled by $k^2$ with $k=1$, 3, 10, 30, 100, and 300, preserving the correlation structure of the flight solution. For each initial uncertainty, a 500 trial Monte Carlo simulation was run with the $\sigma_t$ budget scaled by 1, 2, 4, and 8 (5, 10, 20, and 40 s), injected about the measured umbral midpoint times and declared in $R_k$. All cases are evaluated at the nominal penumbral exit epoch. The percentage of the open-loop $3\sigma$ position error removed at this epoch is plotted in Fig. \ref{fig:capstone_results}.

Starting from the delivered OD covariance, the umbral pair removes 6.3\% of the $3\sigma$ position error at $\sigma_t=5$ s and 0.3\% at 40 s. At a 1.6 km initial uncertainty the reduction is 28.4\% at 5 s, and at 5.33 km it is 60.8\% at 5 s and 8.8\% at 40 s. Above a 16 km initial uncertainty, the reduction at $\sigma_t=5$ s levels off between 70 and 74\%, and at 160 km it is 68.7 to 69.6\% for every $\sigma_t$, since two scalar contacts constrain only two directions of the six-dimensional state. The median NEES lies between 4.3 and 6.0 across all cases, with 3 to 8\% of trials above 12.59.

Fig. \ref{fig:capstone_results_a} shows the 5.33 km, $\sigma_t=5$ s case as a time history. At the last contact, the $3\sigma$ position error drops from 46.41 km open loop to 18.17 km, and the $3\sigma$ velocity error drops from 6.224 m/s to 2.531 m/s. The mean estimated $3\sigma$ position covariance is 18.26 km, a ratio of sample error to covariance of 1.00, and the mean Mahalanobis distances are 0.93 at umbral entry and 1.57 at umbral exit. The benefit decays through the perilune passage, though process noise is turned off; 10 hours after the last contact, the $3\sigma$ position error is 38.90 km against 42.24 km open loop.

Comparing the Earth and lunar eclipse examples, two lunar umbral contacts remove 60.8\% of the $3\sigma$ position error from a 5.33 km initial uncertainty, while four terrestrial contacts remove at most 33.3\% from a 346 km initial uncertainty. This is the combined effect of a tighter event-time uncertainty and a geometry roughly 45 times closer to the occulting body, with two contacts whose constraint directions are well separated. In both cases, the amount of information available cannot compete with a ground station track. The eclipse observables becomes useful once the navigation solution has degraded to the order of kilometers for CAPSTONE and the order of 100 km for the $L_1$ halo, still reasonable orders of magnitude given the size of these dynamical structures in cislunar space. These observables could provide valuable uncertainty reduction in missions relying on autonomous navigation solutions with relaxed uncertainty requirements.

\subsection{Information Retention}\label{sec:retention}
The results above give the uncertainty reduction at the last contact and 10 hours later. To quantify how long the information gained from the eclipse measurements remains relevant, the UKF mean and covariance of the two time-history cases in Figs. \ref{fig:l1_results_a} and \ref{fig:capstone_results_a} are propagated well past the last contact, both with and without the eclipse updates. The filter starts from the a priori mean at the initial filter epoch with the a priori covariance of each case, without sampling initial states, processes the contacts at their nominal times, and propagates the sigma points through the ephemeris model of Eq. \ref{eq:ephem}. The Monte Carlo results above support this covariance-only approach, since the sample error matched the filter covariance with ratios of 1.03 for the $L_1$ halo and 1.00 for CAPSTONE at the last contact. The retained reduction is defined as
\begin{equation}
    \varrho(t)=1-\frac{\sigma_f(t)}{\sigma_{ol}(t)}\label{eq:retention}
\end{equation}
where $\sigma_f$ and $\sigma_{ol}$ are the $3\sigma$ position uncertainties of the filtered and open-loop solutions. The time at which $\sigma_f$ grows back to the open-loop value at the last contact is also recorded, since it is the time the eclipse updates add before the position uncertainty returns to where it would have been without them.
\begin{figure}[htbp!]
    \centering
    \includegraphics[width=0.4\textwidth]{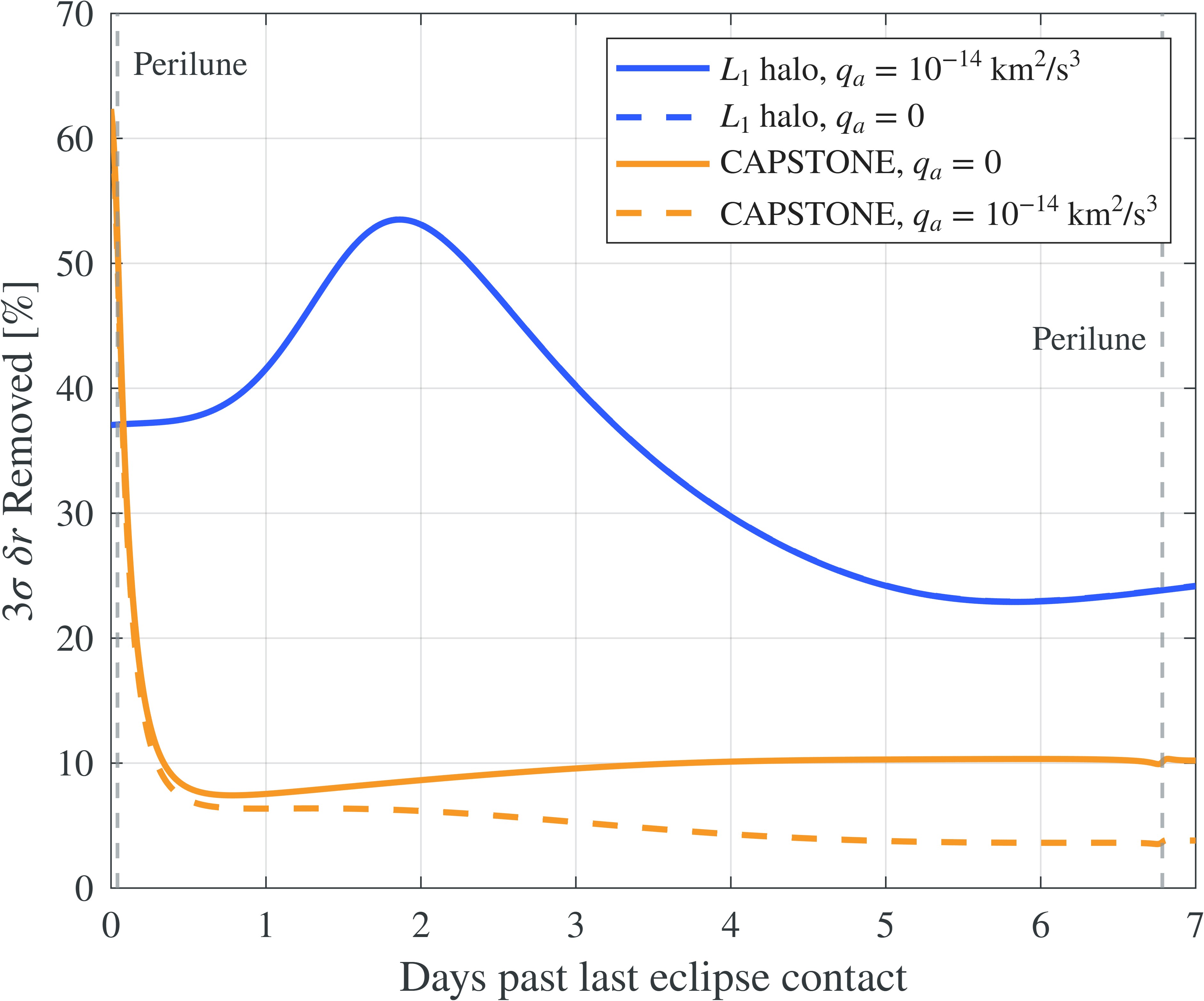}
    \caption{Retained position uncertainty reduction after the last eclipse contact, computed from the UKF covariance. The two $L_1$ halo curves overlap. Perilune labels for CAPSTONE NRHO only.}
    \label{fig:retention}
\end{figure}

Figure \ref{fig:retention} shows $\varrho(t)$ over the seven days after the last contact. Solid lines represent the process noise used for the studies in Figs. \ref{fig:l1_results} and \ref{fig:capstone_results}. For CAPSTONE with $q_a=0$, the covariance gives a reduction of 62.4\% at the last contact, close to the 60.8\% found from the Monte Carlo samples. The perilune passage one hour later amplifies the uncertainty in directions the umbral contacts do not constrain, and the reduction drops to 8.7\% ten hours after the last contact, against 7.9\% from the samples. For the rest of the week it stays between 7.4 and 10.3\%, including the next perilune 6.8 days after the last contact. The filtered $3\sigma$ position uncertainty grows back to the open-loop value at the last contact, 48.5 km, after 22.7 hours. The information retained from processing eclipse transitions is expected to be larger for an eclipse passage occurring post-perilune.

For the $L_1$ halo, the covariance gives a reduction of 37.1\% at the last contact, against 33.4\% from the samples, and the reduction stays between 22.9 and 53.5\% over the week. The filtered $3\sigma$ position uncertainty reaches the open-loop value at the last contact, 1076 km, after 1.83 days. Over the same week the open-loop $3\sigma$ position uncertainty grows from 1076 km to 8767 km as the unstable halo dynamics amplify the initial error. The coasted mean state stays within the Earth range the halo covers in the 24 days before the filter epoch, 317,000 to 390,000 km, for about 20 days, so the week of data shown in Fig. \ref{fig:retention} remains on the halo.

The level of process noise in the true dynamics also affects how long the reduction lasts. Repeating the CAPSTONE case with $q_a=1\times10^{-14}$ km$^2$/s$^3$, under the assumption that the true dynamics carry the same level of unmodeled acceleration, lowers the retained reduction to between 3.5 and 6.4\% from one to seven days after the last contact. It also shortens the time for the filtered uncertainty to return to the open-loop value at the last contact from 22.7 to 16.0 hours. This process noise adds 10 km of $3\sigma$ position uncertainty over the first day along the CAPSTONE trajectory, about a fifth of the 49.4 km filtered uncertainty at that time without process noise. Removing the process noise from the $L_1$ halo case leaves its curve unchanged, because the halo position uncertainty is already hundreds to thousands of kilometers.

\section{Concluding Remarks}\label{sec:6}

This work formulated onboard eclipse detection as an event-based navigation observable for cislunar trajectories that can be used to supplement existing autonomous navigation architectures when available. The time of a penumbral or umbral transition is processed as an implicit zero-measurement on the boundary of a simplified spherical shadow cone, with the event-time uncertainty mapped into the geometric domain through the time sensitivity of the cone function. These observables are ``free" in the sense that no new hardware beyond essential spacecraft equipment is necessary to collect them, with the assumption that a planetary ephemeris kernel is accessible for reference onboard. The methodology was applied to a simulated $L_1$ halo orbit, and data retrieved directly from the CAPSTONE spacecraft operating in its NRHO. The utility in evaluating the eclipse observables is a direct function of the event-time uncertainty $\sigma_t$, which was built from its physical sources: telemetry sampling, the scatter of Earth's atmospheric eclipse-forming layer, and the shape of the occulting body through Earth's oblateness or lunar limb topography. At a 15 s telemetry interval, this gives $\sigma_t$ of 8 to 10 s for the $L_1$ halo Earth eclipse and 5 s for the CAPSTONE lunar umbral contacts.

The eclipse observables were studied with a UKF. As expected, the observables become more useful as the prior navigation solution degrades. For CAPSTONE, the two umbral contacts remove 6.3\% of the $3\sigma$ position error relative to a fresh DSN track solution, but 60.8\% from a 5.33 km initial uncertainty. For the $L_1$ halo, the four contacts remove 33.3\% from a 346 km initial uncertainty. After the last contact, the halo keeps 23 to 54\% of the reduction for at least one week. For CAPSTONE, the reduction falls to about 10\% or less after the perilune that follows the eclipse, and the amount that remains depends on the level of process noise in the dynamics. Eclipse events in cislunar space are sparse, and a single eclipse provides at most two well-separated constraint directions. Processing even a single eclipse transition measurement therefore captures much of the information gain, and subsequent measurements that constrain the same direction give diminishing returns. In a sequential filter, each transition could be processed individually and the best one kept, or all transitions could be accumulated into a batch update.


\section*{Acknowledgments}
The authors would like to thank Matthew Bolliger, Tabitha D'Amato, Alec Forsman, Amanda Herring, Andrew Koehler, and Jeffrey Parker from Advanced Space as well as Hannah Umansky from Terran Orbital for processing and providing the CAPSTONE spacecraft data used in this work.
\bibliography{refs}

@inproceedings{mcelreath2026lazuli,
  author    = {McElreath, James and Carrico, John and Loucks, Mike and Hawkins, Alisa},
  title     = {Preliminary Trajectory Design of the Lazuli Space Observatory},
  booktitle = {AAS/AIAA Astrodynamics Specialist Conference},
  year      = {2026},
  note      = {AAS Paper 26-793}
}

@inproceedings{dichmann2014tess,
  author    = {Dichmann, Donald J. and Parker, Joel J. K. and Williams, Trevor W. and Mendelsohn, Chad R.},
  title     = {Trajectory Design for the Transiting Exoplanet Survey Satellite},
  booktitle = {Proceedings of the 24th International Symposium on Space Flight Dynamics},
  year      = {2014},
  address   = {Laurel, Maryland},
  month     = may,
  note      = {International Symposium on Space Flight Dynamics, May 5--9, 2014},
  number    = {GSFC-E-DAA-TN14089}
}

@article{christian2009review,
  title={Review of options for autonomous cislunar navigation},
  author={Christian, John A and Lightsey, E Glenn},
  journal={Journal of Spacecraft and Rockets},
  volume={46},
  number={5},
  pages={1023--1036},
  year={2009}
}

@book{christian2026fundamentals,
  title={Fundamentals of Spacecraft Optical Navigation},
  author={Christian, John A},
  year={2026},
  publisher={John Wiley \& Sons}
}

@incollection{cheetham2021cislunar,
  title={Cislunar autonomous positioning system technology operations and navigation experiment (Capstone)},
  author={Cheetham, Bradley},
  booktitle={ASCEND 2021},
  pages={4128},
  year={2021}
}

@inproceedings{thompson2022cislunar,
  title={Cislunar navigation technology demonstrations on the CAPSTONE mission},
  author={Thompson, Michael R and Forsman, Alec and Chikine, Sai and Peters, Brian C and Ely, Todd and Sorensen, Dana and Parker, Jeff and Cheetham, Brad},
  booktitle={Proceedings of the 2022 International Technical Meeting of The Institute of Navigation},
  pages={471--484},
  year={2022}
}

@article{sheikh2006spacecraft,
  title={Spacecraft navigation using X-ray pulsars},
  author={Sheikh, Suneel I and Pines, Darryll J and Ray, Paul S and Wood, Kent S and Lovellette, Michael N and Wolff, Michael T},
  journal={Journal of Guidance, Control, and Dynamics},
  volume={29},
  number={1},
  pages={49--63},
  year={2006}
}

@inproceedings{mcelreath2025eclipse,
  title={Utilizing Penumbral Transitions as Onboard Measurement Updates for Cislunar Navigation},
  author={McElreath, James and Down, Ian M and Majji, Manoranjan},
  booktitle={Proceedings of the AAS/AIAA Astrodynamics Specialist Conference},
  year={2025}
}

@article{zimovan2020near,
  title={Near rectilinear halo orbits and nearby higher-period dynamical structures: orbital stability and resonance properties},
  author={Zimovan-Spreen, Emily M and Howell, Kathleen C and Davis, Diane C},
  journal={Celestial Mechanics and Dynamical Astronomy},
  volume={132},
  number={5},
  pages={28},
  year={2020},
  publisher={Springer}
}

@inproceedings{williams2017targeting,
  title={Targeting cislunar near rectilinear halo orbits for human space exploration},
  author={Williams, Jacob and Lee, David E and Whitley, Ryan J and Bokelmann, Kevin A and Davis, Diane C and Berry, Christopher F},
  booktitle={AAS/AIAA Space Flight Mechanics Meeting},
  number={JSC-CN-38615},
  year={2017}
}

@article{fitzgerald2021autonomous,
  title={Autonomous Spacecraft Orbit Determination from Incident Light Intensity via Eclipse Transient Timing},
  author={Fitzgerald, Riley M},
  journal={Journal of Guidance, Control, and Dynamics},
  volume={44},
  number={9},
  pages={1621--1637},
  year={2021},
  publisher={American Institute of Aeronautics and Astronautics}
}

@inproceedings{dewalch2024enhancing,
  title={Enhancing Eclipse Transient Orbit Determination Methods with Statistical Atmospheric Models},
  author={DeWalch, Truman and Fitzgerald, Riley M},
  booktitle={AIAA SCITECH 2024 Forum},
  pages={0429},
  year={2024}
}

@article{park2025bridging,
  title={Bridging ephemeris transition gaps: Quasi-periodic extensions for the Hill Restricted Four-Body Problem in cislunar space},
  author={Park, Beom and Sanaga, Rohith Reddy and Howell, Kathleen C},
  journal={Acta Astronautica},
  volume={233},
  pages={168--182},
  year={2025},
  publisher={Elsevier}
}

@article{park2025frequency,
  title={A frequency-based hierarchy of dynamical models in cislunar space: leveraging periodically and quasi-periodically perturbed models},
  author={Park, Beom and Sanaga, Rohith Reddy and Howell, Kathleen C},
  journal={Celestial Mechanics and Dynamical Astronomy},
  volume={137},
  number={1},
  pages={5},
  year={2025},
  publisher={Springer}
}

@article{sanaga2026systematic,
  title={Systematic Numerical Transitions of Cislunar Trajectories Across Dynamical Models},
  author={Sanaga, Rohith Reddy and Park, Beom and Howell, Kathleen C},
  journal={The Journal of the Astronautical Sciences},
  volume={73},
  number={1},
  pages={10},
  year={2026},
  publisher={Springer}
}

@article{pontani2022optimal,
  title={Optimal low-thrust lunar orbit transfers with shadowing effect using a multiple-arc formulation},
  author={Pontani, Mauro and Corallo, Francesco},
  journal={Acta Astronautica},
  volume={200},
  pages={549--561},
  year={2022},
  publisher={Elsevier}
}

@article{zhang2022comparison,
  title={Comparison of shadow models and their impact on precise orbit determination of BeiDou satellites during eclipsing phases},
  author={Zhang, Yan and Wang, Xiaoya and Xi, Kewei and Li, Zhen},
  journal={Earth, Planets and Space},
  volume={74},
  number={1},
  pages={126},
  year={2022},
  publisher={Springer}
}

@article{zubko2024simplified,
  title={A simplified analytical approach for determining eclipses of satellites occulted by a celestial body},
  author={Zubko, Vladislav and Belyaev, Andrey},
  journal={Acta Astronautica},
  volume={220},
  pages={374--391},
  year={2024},
  publisher={Elsevier}
}

@article{song2016effect,
  title={The effect of the Earth’s oblateness on predicting the shadow conditions of a distant spacecraft: Application to a fictitious lunar explorer},
  author={Song, Young-Joo and Kim, Bang-Yeop},
  journal={Advances in Space Research},
  volume={57},
  number={1},
  pages={355--366},
  year={2016},
  publisher={Elsevier}
}

@article{li2019shadow,
  title={A shadow function model based on perspective projection and atmospheric effect for satellites in eclipse},
  author={Li, Zhen and Ziebart, Marek and Bhattarai, Santosh and Harrison, David},
  journal={Advances in Space Research},
  volume={63},
  number={3},
  pages={1347--1359},
  year={2019},
  publisher={Elsevier}
}

@article{montenbruck2002satellite,
  title={Satellite orbits: models, methods, and applications},
  author={Montenbruck, Oliver and Gill, Eberhard and Lutze, FH},
  journal={Appl. Mech. Rev.},
  volume={55},
  number={2},
  pages={B27--B28},
  year={2002}
}

@article{neta1998satellite,
  title={On satellite umbra/penumbra entry and exit positions},
  author={Neta, Beny and Vallado, David},
  journal={The Journal of the astronautical sciences},
  volume={46},
  number={1},
  pages={91--103},
  year={1998},
  publisher={Springer}
}

@book{crassidis2004optimal,
  title={Optimal estimation of dynamic systems},
  author={Crassidis, John L and Junkins, John L},
  year={2004},
  publisher={Chapman and Hall/CRC}
}

@article{stoker2023aiding,
  title={Aiding Angles-Only Navigation with Illumination Telemetry to Resolve Weak Observability},
  author={Stoker-Spirt, Eric T},
  journal={Journal of Guidance, Control, and Dynamics},
  volume={46},
  number={11},
  pages={2110--2120},
  year={2023},
  publisher={American Institute of Aeronautics and Astronautics}
}

@article{mahalanobis2018generalized,
  title={On the generalized distance in statistics},
  author={Mahalanobis, Prasanta Chandra},
  journal={Sankhy{\=a}: The Indian Journal of Statistics, Series A (2008-)},
  volume={80},
  pages={S1--S7},
  year={2018},
  publisher={JSTOR}
}

@article{bradley2020cislunar,
  title={Cislunar navigation accuracy using optical observations of natural and artificial targets},
  author={Bradley, Nicholas and Olikara, Zubin and Bhaskaran, Shyam and Young, Brian},
  journal={Journal of Spacecraft and Rockets},
  volume={57},
  number={4},
  pages={777--792},
  year={2020},
  publisher={American Institute of Aeronautics and Astronautics}
}

@inproceedings{wan2000unscented,
  title={The unscented Kalman filter for nonlinear estimation},
  author={Wan, Eric A and Van Der Merwe, Rudolph},
  booktitle={Proceedings of the IEEE 2000 adaptive systems for signal processing, communications, and control symposium (Cat. No. 00EX373)},
  pages={153--158},
  year={2000},
  organization={Ieee}
}

@article{herald2014analysis,
  title={Analysis of lunar crater timings, 1842-2011},
  author={Herald, David and Sinnott, Roger W},
  journal={Journal of the British Astronomical Association},
  volume={124},
  pages={247--253},
  year={2014}
}

@article{petit2010iers,
  title={IERS conventions (2010)},
  author={Petit, G{\'e}rard and Luzum, Brian and others},
  year={2010},
  publisher={Bundesamt f{\"u}r Kartographie und Geod{\"a}sie Frankfurt am Main, Germany}
}

@article{smith2010initial,
  title={Initial observations from the lunar orbiter laser altimeter (LOLA)},
  author={Smith, David E and Zuber, Maria T and Neumann, Gregory A and Lemoine, Frank G and Mazarico, Erwan and Torrence, Mark H and McGarry, Jan F and Rowlands, David D and Head III, James W and Duxbury, Thomas H and others},
  journal={Geophysical Research Letters},
  volume={37},
  number={18},
  year={2010},
  publisher={Wiley Online Library}
}

@book{widrow2008quantization,
  title={Quantization noise},
  author={Widrow, Bernard and Koll{\'a}r, Istv{\'a}n},
  volume={2},
  number={5},
  year={2008},
  publisher={Cambridge University Press Cambridge}
}

@misc{espenaklimb,
  author       = {Espenak, Fred},
  title        = {The Lunar Limb Profile and Eclipse Predictions},
  howpublished = {NASA Goddard Space Flight Center Eclipse Web Site, \url{https://eclipse.gsfc.nasa.gov/SEhelp/limb.html}},
  year         = {n.d.},
  note         = {Accessed September 2026}
}

\end{document}